\documentclass[12pt,preprint]{aastex}

\def\arcsecpoint{$''\!.$}
\def\deg{$^{\rm o}$}
\def\ltsim{\raisebox{-.5ex}{$\;\stackrel{<}{\sim}\;$}}

\shortauthors{Kraemer et al.}
\shorttitle{UV Absorbers in NGC~4151}

\begin{document}

\title{Simultaneous Ultraviolet and X-ray Observations of the Seyfert Galaxy NGC
4151. II. Physical Conditions in the UV Absorbers\altaffilmark{1}}

\author{S.B. Kraemer\altaffilmark{2,3}, 
D.M. Crenshaw\altaffilmark{4},
J.R. Gabel\altaffilmark{5},
G.A. Kriss\altaffilmark{6},
H. Netzer\altaffilmark{7},
B.M. Peterson\altaffilmark{8},
I.M. George\altaffilmark{9,10},
T.R. Gull\altaffilmark{3},
J.B. Hutchings\altaffilmark{11},
R.F. Mushotzky\altaffilmark{10},
\& T.J. Turner\altaffilmark{9,10}}

\altaffiltext{1}{Based on observations made with the NASA/ESA {\it Hubble Space 
Telescope}, obtained at the Space Telescope Science Institute, which is 
operated by the Association of Universities for Research in Astronomy, Inc., 
under NASA contract NAS 5-26555; these observations are associated with 
proposal GO-9272. Also based on observations made with the NASA-CNES-CSA {\it
Far Ultraviolet Spectroscopic Explorer}. {\it FUSE} is operated for NASA by
Johns Hopkins University under NASA contract NAS5-32985.}

\altaffiltext{2}{Institute for Astrophysics and Computational Sciences,
Department of Physics, The Catholic University of America, Washington, DC
20064; kraemer@yancey.gsfc.nasa.gov}

\altaffiltext{3}{Exploration of the Universe Division, Code 667, NASA's
Goddard Space Flight Center, Greenbelt, MD 20771}

\altaffiltext{4}{Department of Physics and Astronomy, Georgia State University,
Atlanta, GA 30303}

\altaffiltext{5}{University of Colorado, CASA, UCB 389, Boulder, CO 80309-0389}

\altaffiltext{6}{Space Telescope Science Institute, 3800 San Martin Dr. Baltimore, Md 21218}

\altaffiltext{7}{School of Physics and Astronomy, Raymond and Beverly Sackler Faculty of
Exact Sciences, Tel-Aviv University, Tel-Aviv 69978, Israel}

\altaffiltext{8}{Department of Astronomy, The Ohio State University, 140 W. 18th Ave.,
Columbus, OH 43210-1173}

\altaffiltext{9}{Physics Department, University of Maryland, Baltimore
County, Baltimore, MD 21250}

\altaffiltext{10}{Exploration of the Universe Division, Code 662, 
NASA Goddard Space Flight Center, Greenbelt, MD 20771}

\altaffiltext{11}{Dominion Astrophysical Observatory, National Research Council of Canada, 
Victoria, BC V8X4M6, Canada}

\begin{abstract}

We present a detailed analysis, including photoionization modeling, of the intrinsic absorption in
the Seyfert 1 galaxy NGC 4151 using ultraviolet (UV) spectra from the
{\it Hubble Space Telescope}/Space Telescope Imaging Spectrograph and
the {\it Far Ultraviolet Spectrographic Explorer} obtained 2002 May as part of
a set of contemporaneous observations that included
{\it Chandra}/High Energy Transmission Grating Spectrometer spectra. 
In our analysis of the {\it Chandra} spectra, we determined 
that the X-ray absorption was dominated by two components: a high-ionization absorber,
revealed by the presence of H-like and He-like lines of Mg, Si, and S, and 
a lower-ionization absorber, in which inner-shell
absorption lines from lower-ionization species of these elements
formed. We identified the latter as the source of the saturated UV lines of O~VI, C~IV, and N~V
associated with the absorption feature at a radial velocity of $\sim$ $-$ 500 km s$^{-1}$, which we referred to as component
D$+$E.
In the present work, we have derived tighter constrains on the the line-of-sight covering factors,
densities, and radial distances of the absorbers. We confirm the presence of the 3 sub-components of D$+$E
described in our previous paper, with line-of-sight covering factors ($C_{los}$) ranging from $\sim$ 0.5 -- 0.9,
and find evidence for a fourth component, D$+$Ed characterized by low ionization
and a $C_{los}$ $\sim$ 0.2. The complexity of the UV absorption in NGC 4151 may be a consequence of the fact that we are
viewing the black hole/accretion disk system at a relatively high inclination and, therefore, may be
detecting the densest part of the flow. Our deconvolution of the underlying C~IV emission 
indicates that D$+$E must lie outside the Intermediate Line Region, hence at a radial distance
of $\sim$ 0.1 pc. We find that the Equivalent Widths (EWs) of the low-ionization lines associated
with D$+$E varied over the period from 1999 July to 2002 May. Although over part of this time, the variations were correlated
with changes in the UV continuum, the drop in the EWs of these lines between 2001 April and 2002 May are
suggestive of bulk motion of gas out of our line-of-sight. Over this period, $C_{los}$ for the low-ionization absorption lines
dropped from $\sim$ 0.7 to $\sim$ 0.2. If these lines from these two epochs arose in the same sub-component, the transverse
velocity of the gas is $\approx$ 2100 km s$^{-1}$. This is similar to the constraint on transverse velocity derived from the
drop in the X-ray absorbing column between 2000 March and 2002 May. Transverse velocities of this order are consistent
with an origin in a rotating disk, at the roughly radial distance we derived for D$+$E. As we suggested in our previous
study, it is likely that the absorption arises in a disk-driven wind.

\end{abstract}

~~~~~

\section{Introduction}

Understanding the nature of mass outflow in active galactic nuclei (AGN) has been an area
of extensive recent research activity. The UV and X-ray spectral properties of such outflows
 and their overall physical properties have been detailed by
Crenshaw, Kraemer, \& George (2003). To summarize, the ubiquity of blue-shifted absorption among Seyfert galaxies indicates that the 
global covering factors of the absorbers must be high. The total mass in the outflows can exceed the
mass of gas in the broad line region (BLR). Furthermore, the mass outflow rates are on the same
order as the accretion rates required to fuel the supermassive black holes that power AGN.
Hence, determining the nature of these outflows is critical to understanding
the structure and evolution of AGN. However the origin of the outflows, e.g. an accretion disk (Konigl \& Kartje
1994) or the putative torus (Krolik \& Kriss 1995, 2001), and the physical means by which
they are accelerated, e.g. thermal winds (Krolik \& Begelman 1986), radiation pressure (Murray et al.
1995), or magneto-hydrodynamic (MHD) flows (Bottorff, Korista, \& Shlosman 2000), remain uncertain. 

NGC 4151 ($cz =  995$ km s$^{-1}$) was the first Seyfert galaxy to show intrinsic absorption. Oke \& Sargent (1968) 
found non-stellar He~I $\lambda$3889 and Anderson \& Kraft (1969) detected H$\beta$ and H$\gamma$ self-absorption.
These lines were blue-shifted with respect to the host galaxy, indicating mass outflow. Observations in the UV 
obtained with the {\it IUE} (Boksenberg et al. 1978) and far-UV using the Hopkins UltraViolet Telesocope (HUT; Kriss et al., 1992, 1995)
revealed absorption lines from a wide range
of ionization states, including fine-structure and metastable lines. The intrinsic lines were found
to vary in response to the UV continuum (Bromage et al. 1985), suggesting ionic column density changes resulting
from changes in the ionization state of the absorbers. Monitoring of NGC 4151 in the far-UV with HUT (Kriss et al. 1995)
and the {\it Orbiting and Retrievable Far and Extreme Ultraviolet Spectrometers-Shuttle Pallet
Satellite II} mission (Espey et al. 1998) also showed evidence for short timescale variations in absorption due to changes
in ionization. Early X-ray spectra
revealed the presence of a large column of absorbing gas (Barr et al. 1977; Holt et al.
1980), which subsequent observations revealed to be ionized (Yaqoob, Warwick,
\& Pounds 1989; George et al. 1998). As with the UV absorption, the X-ray absorption is
highly variable (George et al. 1998; and references therein), but the changes are not always
correlated with the X-ray continuum flux and may indicate transverse motion of the gas across the line of sight (Weaver et al. 1994).
  
In the {\it Hubble Space Telescope/HST} survey of intrinsic UV absorption in Seyfert galaxies (Crenshaw et al.
1999), NGC 4151 stood out as a unique object. It is one of the few objects whose spectra that show Mg~II absorption, and
it is the only one that showed strong metastable C~III* $\lambda$ 1175 absorption, until recent
detection of a weak C~III* absorption line in NGC 3783 (Gabel et al. 2003). One possible reason for the
complexity of the intrinsic absorption in NGC 4151 is the fact that we are viewing the black hole/accretion
disk system at a relatively high inclination of $\sim$ 45\deg~with respect to the accretion disk (Evans et al. 1993; Das et al. 2005), and 
are likely detecting material close to the densest part of the outflow. The implication is that this 
material cannot be detected in absorption if the AGN is viewed more face-on. 

The first high-resolution ($\sim$ 15 km s$^{-1}$) UV spectra
obtained by Weymann et al. (1997) using the Goddard High Resolution Spectrograph (GHRS) aboard {\it HST} revealed
six major kinematic components of C~IV and Mg~II absorption, which were stable over the
period 1992--1996.  Following this, we obtained an {\it HST}/Space Telescope Imaging Spectrograph (STIS) 
GTO spectrum in 1999 July (Kraemer et al. 2001; hereafter K01). These spectra revealed that the kinematic components detected
by Weymann et al. were still present, although the strongest kinematic components, D and E using the nomenclature of
Weymann et al.,  were broad and we were unable to separate them. 
Also, we found complex  
line-of sight covering factors\footnote{Here, we generalize
partial covering factor $C_{los}$ to include either unocculted regions in our direct view
to the nuclear emission sources, or light that is filling
in the absorption troughs due to an extended scatterer.} ($C_{los}$) 
in the broad D+E component.
Based on the relative depths of individual members of doublets and
multiplets, different ions in this component were found to have different
covering factors:  
the strong high-ionization \ion{C}{4} and \ion{N}{5} doublets 
had $C_{los} \approx$ 0.9, the \ion{Si}{4} and \ion{Mg}{2} doublets yielded $C_{los} \approx $0.85
and 0.62, respectively, and the many \ion{Fe}{2} metastable lines 
indicated even lower covering factors. From our photoionization modeling analysis, we argued that
the complex, low ionization absorption arose in dense gas, within 0.03 pc of the central source.
The presence of these low ionization lines was a consequence of the relatively low continuum
state during this observation, as originally suggested by Bromage et al. (1985). However, the
lack of contemporaneous X-ray observations prevented us from fully constraining the physical
state and geometry of the circumnuclear absorbers.

In order to better characterize the absorbers, we obtained contemporaneous {\it FUSE}, {\it HST}/STIS, 
and {\it Chandra}/High Energy Transmission Grating Spectrometer(HETGS) spectra in 2002 May (see next section). 
In Kraemer et al. 2005 (hereafter Paper I),
we presented our analysis of the {\it Chandra} observations. To summarize,
we were able to model the intrinsic X-ray absorption using five distinct components (see Table 1). 
The H-like Mg, Si, and
S lines were consistent with the presence of a highly ionized component, X-High, which was not detected
in the UV spectra. 
The bulk of the soft X-ray absorption is associated with the 
UV kinematic component D$+$E. Based on our preliminary analysis of the UV absorption,
we found evidence for three subcomponents associated with D$+$E. The saturated O~VI, N~V, and C~IV lines have $\sim$ 10\% 
residual flux in the deepest part of their troughs. We identified this component, D$+$Ea, as the source of the inner-shell lines of
S, Si and Mg detected in the HETG spectrum.
The \ion{C}{3}*~$\lambda$1175 absorption, arising from the $2s2p~^{3}P$ metastable levels, over this range of velocity 
shows a higher residual flux ($\sim$ 20\%), hence, we added a 
second component, D$+$Eb. Finally, the \ion{P}{5} doublet
indicated the presence of a third component, D$+$Ec, with a much lower covering factor ($C_{los}  \approx 0.5$). 
A fifth component of 
X-ray absorption was identified with
UV kinematic component E$'$.  

In Paper I, we also analyzed the 2000 March 3 {\it Chandra}/HETGS observation of NGC 4151. We found that the
2-10 keV continuum flux was a factor of $\sim$ 2 lower than in 2002 and that the soft X-ray continuum was
much more heavily absorbed. Although higher soft X-ray opacity would be a consequence of the drop in
ionization of the absorbers, due to the weaker ionizing flux, in order to fit the spectrum we required
an additional column of gas, which we associated with component D$+$Ea, in which case it must
cover the continuum source and the Broad Line Region (BLR). This column must have moved out of 
our line-of-sight in the time between the two observations, which requires a transverse velocity\footnote{in Paper I, we had
used the radius of the broad line region to estimate v$_T$, while, here, we correctly use the diameter.}  
$v_{T} > 2500$ km s$^{-1}$.

Based on our photoionization models, we found that X-High could not be accelerated via radiation pressure or as
part of a thermal wind. Although the other components could be radiatively driven, the overlap
in velocities of X-High and the components of D$+$E suggests a common mechanism. We suggested that the most 
likely means for driving the outflow is by an MHD wind arising from the accretion disk. Interestingly, a disk 
origin would provide a natural mechanism for the observed
transverse bulk motion: i.e., rotation.

In the present study, we have used the parameters determined in Paper I as the initial
model inputs, subject to modest adjustments in order to refine the fit to the UV absorption, with the main
goal being to more tightly constrain the radial distances of the absorbers. We also track the changes
in the UV absorption over the period from 1997-2002 to examine the effects of the response to the 
ionizing continuum and/or bulk motion.

\section{Observations}

We obtained new UV and X-ray spectra of the nucleus of NGC~4151 with the {\it
HST}/STIS echelle gratings on 2002 May 8, {\it FUSE} on 2002 May 28, and {\it
Chandra} on 2002 May 7 -- May 11. The {\it Chandra} observations are
described in Paper I. For comparison, we retrieved all previous STIS echelle
and {\it FUSE} spectra of NGC~4151 from the Multimission Archive at the Space
Telescope Science Institute (MAST); the details of the observations are
listed in Tables 2 and 3. Most of the STIS echelle observations
come from a STIS key project (J. Hutchings, PI); only the first set of these,
from 1999 July 19, have been published (Crenshaw et al. 2000; K01). The STIS 
echelle observations used a 0\arcsecpoint2 $\times$
0\arcsecpoint2 aperture, whereas the {\it FUSE} observations used a 30$''$
$\times$ 30$''$ (``LWRS'') aperture; the larger aperture has little effect
except to include more emission from the narrow-line region (NLR). Tables 2 and
3 show that we have 6 epochs of STIS E140M observations, 4 epochs of {\it FUSE}
observations, and a smaller number of STIS echelle observations in the longer
wavelength regions. The velocity resolution of the UV spectra ranges from 7 to
15 km s$^{-1}$ (FWHM).

We reduced the STIS spectra using the IDL software developed at NASA's 
Goddard Space Flight Center for the STIS Instrument Definition Team.
The data reduction included a procedure to remove the background light from 
each order using a scattered light model. The individual orders in each 
echelle spectrum were spliced together in the regions of overlap.
We processed the {\it FUSE} data with version 2.2.3 of the standard
calibration 
pipeline, CALFUSE, which extracts spectra for each of the eight combinations 
of channels (SiC1, SiC2, LiF1, LiF2) and detector segments (A and B).  
The channel/segments with overlapping wavelength 
coverage with the LiF1a spectrum were scaled to match its flux.
All spectra were then co-added by weighting each channel/segment by 
its effective area, using the effective area versus wavelength 
functions given by Blair et al. (2000).
We compared Galactic lines in the 
{\it FUSE} and STIS spectra to place them on the same wavelength scale.
Due to non-linear offsets in the {\it FUSE} wavelengths, 
local shifts were measured and applied individually to
small spectral regions containing the intrinsic absorption features.
Mean residual fluxes measured in the cores of 
saturated Galactic lines are consistent with zero within the noise (i.e., standard 
deviation of the fluxes) in the troughs of these lines, indicating accurate background removal.

To place the high-resolution spectra into context, we retrieved for comparison all of the
previous low-resolution spectra of NGC~4151 that contain the 1350 \AA\
continuum region. This includes spectra obtained by the {\it International
Ultraviolet Explorer} ({\it IUE}), {\it the Hopkins Ultraviolet Telescope }
({\it HUT}), and the Faint Object Spectrograph (FOS) on {\it HST}. 
We measured continuum fluxes by averaging the points in a bin centered 
at 1350 \AA\ (observed frame) with a width of 30 \AA, and determined the 
one-sigma flux errors from the standard deviations. For the {\it IUE} spectra, 
this method overestimates the uncertainties (Clavel et al. 1991), and we therefore 
scaled them by a factor of 0.5 so that on average observations taken on the same 
day agree to within the quoted errors.

Figure 1 gives the UV continuum light curve of NGC 4151, which spans a
$\sim$24-year time period. The light curve shows the long-term decline and rise
of the continuum flux during the span of {\it IUE} observations, and an apparent
long-term decline in recent years (although this trend is highly undersampled, since the loss of {\it IUE}).
Superimposed on the long-term trends are more rapid variations captured by
intensive monitoring, such as the continuous monitoring campaign around JD
2,449,300 (Crenshaw et al. 1996). The last six points are STIS echelle
observations obtained
at moderate to low flux states compared to NGC~4151's previous behavior.
Several STIS observations were obtained at very low continuum fluxes, including
an all-time low of F$_{\lambda}$(1350) $=$ 1.4 ($\pm$0.6) x 
10$^{-14}$ ergs s$^{-1}$ cm$^{-2}$ \AA$^{-1}$ on 2001 April 14 (JD 2,452,014).
The broad emission lines tend to get very weak at these low continuum states,
but faint broad wings in the C~IV profiles indicate that they never fully
disappear. We note that the first and last STIS echelle observations
were obtained at very similar continuum flux levels, which is important for our
later discussions. 

\section{Absorption-Line Variations}

In Figure 2, we show the STIS echelle spectra of NGC~4151 in the C~IV region,
along with a STIS G140M spectrum ($\lambda$/$\Delta\lambda$ $\approx$ 15,000)
obtained in a high state on 1997 May 15 (Hutchings et al. 1998). In K01,
we presented detailed analyses of the kinematic absorption components in the 1999 spectra; we 
reiterate some of our principal results here.
Using the Weymann et al. (1997) nomenclature, component B arises from our Galaxy
and components F and F$'$ (in the red wing of F) arise in the interstellar
medium and/or halo of NGC~4151. We were not able to separate Weymann et al.'s D
and E components, and therefore treated them as a single (``D$+$E'') kinematic
component. Weymann et al. also identified a narrow absorption feature (E$'$) in
the low-ionization lines that was not detected in the high-ionization lines.
This component can be seen in Figure 2 in the red wings of the broad D$+$E
components for Si~II $\lambda$1527 and Si~II* $\lambda$1533. We also
identified a broad transient component (D$'$) in C~IV and N~V that appears
blueward of D$+$E. In K01, we give radial velocity centroids
and widths for all of the kinematic components in the 1999 STIS echelle
spectra. We also give models of each intrinsic component based on its measured
ionic column densities. A model is characterized by an ionization parameter
\footnote{
$U = Q/4\pi~r^{2}~c~n_{H}$
where $r$ is the radial distance of the absorber and $Q = \int_{13.6 eV}^{\infty}(L_{\nu}/h\nu)~d\nu$,
or the number of ionizing photons s$^{-1}$ emitted by a source of luminosity
$L_{\nu}$} ($U$), hydrogen column density ($N_H$ $=$ $N_{H^{\rm 0}}$ $+$ $N_{H^{\rm +}}$, in units of cm$^{-2}$), and hydrogen number 
density ($n_H$, in units of cm$^{-3}$); the
latter was determined from numerous absorption lines arising from fine-structure
levels above ground as well as metastable levels in the case of D$+$E.
 
In this paper, we concentrate on the components that are intrinsic to
the AGN in NGC~4151 and are clearly identified in all spectra: A, C,
D$+$E, and E$'$. The 2002 observations have the highest signal-to-noise ratios
of all of the STIS echelle spectra due to the moderate flux levels and long
exposure times, and we use these observations to identify a few more lines than
were originally detected in the 1999 spectra. For each of the above kinematic
components, we have also measured the equivalent width (EW) and
radial-velocity centroid of each detected line, since these measurements
provide the simplest and most sensitive method of characterizing variations in
the absorption lines. We note that equivalent-width variations can result from
changes in ionic column density and/or covering factor of the background
emission; we explore these issues further in subsequent sections.  No
significant changes were detected in radial-velocity centroids, except as
discussed below for component D$+$E.

For component A, we identified lines in the 1999 spectra (including several fine
structure lines) from H~I, C~II, C~IV, Mg~II, Si~II, Si~III, Si~IV, and
Fe~II. In the 2002 spectra, we also detect Al II $\lambda$1671 and Al III
$\lambda\lambda$1855, 1863, as well as C III~$\lambda$977 and N III~$\lambda$989 in the
{\it FUSE} spectrum.  In K01, we noted a decrease
in the EWs of the Si~II and Mg~II lines between GHRS observations on 1994
October 28 and 1996 November 3, which we attributed to transverse motion of
low-ionization gas across the BLR. Figure 3 gives the STIS EWs
of three (medium, low, and high ionization) lines from component A as a
function of time, along with the continuum light curve. There is no
strong evidence for variability in any of the lines since the 1999 observations,
even if we consider only the points with small errors, corresponding to the
highest continuum fluxes (i.e., the first, fourth, and sixth points). Thus, our
model for component A in 1999 ($\log U = -2.92$, $N_H =$ 10$^{18.1}$ cm$^{-2}$,
$n_H \approx$ 10$^{2.6}$ cm$^{-3}$) is still valid for the later STIS observations.

For component C, we detect the same lines in the 2002 spectra that we identified
in the 1999 spectra, from H~I, C~II, C~IV, Mg~II, Si~II, Si~III, and Si~IV; we
detected no fine-structure lines. Lines from C III and N~III are detected in the
{\it FUSE} spectrum.  Figure 4 shows that there is
no evidence for EW variations in component C, as was the case for observations
prior to 1999. Our previous model for this component ($\log U =-2.92$, $N_H =$ 
10$^{18}$ cm$^{-2}$, $n_H\approx$ 10 cm$^{-3}$) is still valid for all
observations.

For the broad D$+$E component, we detect most of the lines in the 2002 spectra
that we saw in the 1999 spectra, which are due to H~I, C~II, C~III, C~IV, N~V,
Mg~II, Al~III, Si~II, Si~III, Si~IV, S~II and Fe~II. However, we do not detect
O~I and Ni~II in the 2002 spectra. 
The {\it FUSE} spectrum exhibits lines from additional ions, S III, S IV, S~VI,
C III, N III, P V, and O VI.  In both the 1999 and 2002 spectra,
we detect the metastable C~III $\lambda$1175 lines and Fe~II lines (weakly present in 2002) from levels
as high as 4.1 eV above the ground state, indicating high densities. As shown in
Figure 2, the high-ionization C~IV line does not show strong EW
variations, although there is some evidence for slight variations in the
residual emission in the troughs. The lack of strong variations in the
high-ionization lines (C~IV, N~V, Si~IV) is likely due to the fact that they are
highly saturated. The low-ionization species (e.g., Si II, Fe~II) show strong
variations in their EWs and we have attributed the variations between 1997 and
1999 to changes in the ionizing continuum (K01). However,
comparison of the 1999 and 2002 observations reveal a more complicated story;
the continuum fluxes are about the same, but the low-ionization lines have
decreased greatly in EW, as can be seen in the Si II~lines in Figure 2.

Figure 5 shows the equivalent width variations of one high-ionization and
two low-ionization (one fine-structure) lines from D$+$E. Given the errors,
there is no strong evidence for variations in C~IV or other high-ionization
lines. The equivalent widths of the low-ionization lines are strongly
anti-correlated with continuum flux for the first five observations, indicating
that ionizing continuum variations are the dominant source of variability.
However, the last observation, in 2002, shows a strong drop in the EWs, despite
a similar continuum level as the first observation in 1999. This holds true
for all of the low-ionization species that have creation ionization potentials
less than 19 eV (corresponding to Al III), including the metastable Fe II lines.
This decrease in EW could be due to a change in total column density ($N_H$)
and/or covering factor for the low-ionization gas; either one requires
transverse motion across the BLR over a period of time $<$ 13 months. 
The low ionization lines also show a shift in their radial-velocity centroids
between 1999 and 2002, from $-$490 km s$^{-1}$ to $-$280 km s$^{-1}$ relative to
the systemic velocity of the host galaxy, as demonstrated by the Si II lines in
Figure 2. The most likely source of this shift is that D$+$E is composed of
several kinematic subcomponents that are blended together, and one or more of
the subcomponents at higher velocities have diminished or even disappeared in
the 2002 spectra. This possibility is discussed in more detail later, after we
present new photoionization models of the D$+$E component.

For component E$'$, we identified only low-ionization (including fine-structure)
lines in the 1999 spectra, from H~I, C~II, O~I, Mg~II, Si~II, and Fe~II. In the
2002 spectra, we also detect Al II $\lambda$1671, Al III $\lambda\lambda$1855,
1863, Ni II $\lambda$1371, Mn II $\lambda$2576, S II, and metastable Fe~II lines from levels $\ltsim$ 1.1 eV above ground (e.g.,
UV multiplets 62 and 63). 
We had predicted the latter would be present, based on
our estimates of the density of this component ($n_H \geq$ 10$^{6}$ cm$^{-3}$).
The {\it FUSE} spectrum also shows lines from S III, S IV, Fe III and metastable 
Si III (6.2 eV above ground).  
In K01, we found that the strengths of the Si~II and Mg~II
lines increased between 1994 October 28 and 1999 July 19. Figure 6 shows that
the EWs of the low-ionization lines in this component have continued to
increase. Unfortunately, the signal-to-noise for spectra obtained in low flux states
does not allow accurate measurement of the lines from this component. However,
given the similar continuum fluxes for the first and last echelle observations,
it is clear that the source of the absorption variations cannot be ionization
changes alone, indicating that transverse motion is important. We present new
photoionization models of E$'$ in the next section.

In the 1999 spectra, component D$^{'}$ was identified via the presence of broad
N~V, C~IV, and Si~IV absorption (K01). In Figure 7, we compare the 1999 and 2002 STIS echelle
spectra in the region near C~IV. In 2002, there still appears to be a flattening of the
profile in the velocities spanned by D$^{'}$, although this is less evident in other
epochs (see Figure 2). The N~V profile shows a similar effect. However, it is also clear that the
underlying emission-line profile was quite different during these two observations, which
hampers a direct comparison of the absorption. Therefore, while we conclude
that D$^{'}$ was likely present in 2002, we cannot accurately determine ionic column densities due
to the difficulty in deconvolving the effects of changes
in the underlying emission and depth of the absorption.
 
\section{Intrinsic UV Absorption Measurements from {\it FUSE} and STIS Observations: 
2002 May Epoch}

\subsection{Component D+E}

As discussed in Paper I, there is evidence for at least three different regions comprising the D+E absorption.
Here, we detail how covering factor constraints derived from the 2002 May {\it FUSE} and
STIS spectra were used to isolate the different regions. First, many lines, including 
the doublets of the lithium-like CNO ions and Lyman lines
show deep troughs from $-$200 to $-$700 km s$^{-1}$, with some filling by unocculted NLR emission (see Figure 8a).
All these lines have residual normalized fluxes $\lesssim 0.1$, giving a lower limit on $C_{los}$ $\sim$ 0.9, consistent with K01.
A number of other lines, including
\ion{N}{3}~$\lambda$ 991, and \ion{S}{6}~$\lambda\lambda$933,944, show similar residual fluxes (see Figures 8a and 8b), but the deepest
parts of their profiles range from 0 to $-$500 km s$^{-1}$. Each of these lines
are heavily saturated, with lower limits to their column densities $\gtrsim$ $10^{15}$ cm$^{-2}$. 
The offset in velocity may be evidence of the different contributions from components D and E to the 
composite profile, in which case it appears that E may be more heavily saturated. However, these components
strongly blend together to form a smooth profile, therefore it is impossible to deconvolve them. 
Although \ion{C}{3}~$\lambda$977 spans a broader
range in velocity than \ion{N}{3} (See Figure 8b), it appears that the blue-wing is affected by the deep absorption
from component C, hence the absorption profile from D$+$E for these two ions is likely similar. 
Based on our analysis of the X-ray absorption (Paper I), we 
associate all of these absorption lines with 
the sub-component D$+$Ea, which is the source of the broad-band soft X-ray absorption.

The strong \ion{C}{3}*~$\lambda$1175 absorption complex shows a velocity structure similar to \ion{S}{6}, but
a smaller residual flux ($\gtrsim$ 0.2), which we
cited in Paper I as evidence for the existence of sub-component D$+$Eb.
The \ion{C}{3}* complex is comprised of six transitions arising from 
the three metastable levels (J=0,1,2) at $E \approx$ 6.2 eV above
the ground state.  
The populations of the metastable levels, and thus strength of the feature,
depend strongly on electron temperature and density (we will revisit this
in Section 5).  The \ion{Si}{4}~$\lambda\lambda$1394,1403 doublet, which also
appears to be saturated, shows a residual flux that ranges between that of the lithium-like CNO lines and the
lines associated with D$+$Eb (see Figure 8c), which suggests that there are contributions
from both sub-components.
  
    The \ion{P}{5} $\lambda$1118,1128 doublet exhibits unsaturated, 
unblended absorption in component D+E, providing
a good diagnostic for covering factor and ionic column density.
It also has negligible emission-line flux underlying the absorption, and thus
will not be affected by complex background emission geometry (see Ganguly
et al. 1999; Gabel et al. 2003).   
We used a $\chi^2$ analysis to derive the covering factor and optical
depth $\tau$ from the \ion{P}{5} doublet, 
comparing the observed intensity for each line with the absorption equation 
that gives normalized intensity as a function of these parameters:
\begin{equation}
I = C e^{-\tau} + (1 - C).
\end{equation}
This method is needed to assess uncertainties in the doublet solution, since
spectral noise is known to bias the solution to underestimates of $C_{los}$ and overestimates 
of $\tau$ (see Appendix of Gabel et al. 2005a).
For 1$\sigma$ uncertainties in the parameters, we adopted 
the maximum offsets from the best fit values 
that give a total $\chi^2 = 2$ (one per doublet member).
Solutions to velocity-resolved $C_{los}$ and $\tau$ 
with 1$\sigma$ error bars are plotted in Figure 9, showing $C_{los} \approx 0.5$
consistently for the independent solutions in each velocity bin.
This is in contrast with the lines we have associated with D$+$Ea and D$+$Eb, 
which clearly have a larger covering factor based
on their low residual fluxes. Therefore, we required a third component, D$+$Ec.

To further test uncertainties 
in the \ion{P}{5} $C_{los} - \tau$ solutions, we did similar analysis with 
\ion{Fe}{2} lines in component B, which is known to fully occult 
the AGN emission ($C_{los} = 1$) due to its location in the Milky Way.
We fit lines having a range of oscillator strengths in relatively 
uncontaminated regions of the {\it FUSE} spectrum 
(\ion{Fe}{2}~$\lambda$1055,1112,1122,1143) to insure similar 
uncertainties due to noise and normalization as the \ion{P}{5} fitting.
$\chi^2$ analyses of different combinations of these lines all give $C_{los} = 1$ within 
1$\sigma$ uncertainties, thereby verifying the solution and 
error estimates for \ion{P}{5}.

Many other UV lines are expected to have strong contributions from D$+$Ec, particularly
given the relatively low abundance of phosphorus compared to elements such as carbon, nitrogen,
and oxygen. However, it is effectively impossible to detect these lines if there are significant
columns of these ions in the two other subcomponents of D$+$E, due to their higher
covering factors. This is particularly true for D$+$Eb, since the X-ray photoionization models predict
ionic columns similar to those in D$+$Ec (Paper I). The effect of overlapping components can be
seen in the case of the \ion{C}{2}~$\lambda$1335,  
\ion{C}{2}*~$\lambda$1336, and \ion{S}{4}*~$\lambda$1073 lines (see Figure 10). These
have line depths somewhat greater than the unocculted flux level implied by the \ion{P}{5} 
covering factor, with $C_{los} \sim 0.6 - 0.8$, which is consistent with saturation of both lines in D$+$Ec and
a potentially strong contribution from D$+$Eb. Lower limits to the column densities of 
these ions, derived by assuming $C_{los} = 0.8$, are given in Table 4 .

The relative contributions to the \ion{Mg}{2}~$\lambda\lambda$2795,2803 doublet are even more difficult
to deconvolve, since
the intrinsic emission-line profile is difficult to determine due to the broad D+E 
absorption and the presence of underlying broad Fe~II emission (see Verner et al. 2003), but is
consistent with a low covering factor.
The lines from two fine-structure levels of
\ion{S}{3}, \ion{S}{3} *$\lambda$1015 (J=1, $E_{ll}=$299 cm$^{-1}$; the
latter is the energy, above ground, of an excited state, hence the lower-level energy of the
observed transition) and \ion{S}{3} **$\lambda$1021 (J=2, $E_{ll}=$833
cm$^{-1}$), are present in D$+$E.
There is only an upper limit on the ground-state column density 
(see Figure 8d), while the lines from the two excited levels exhibit
shallow features. We measured limits on the column densities for both
D$+$Eb and D$+$Ec by using $C_{los} =0.8$ and the optical depth profile of
S~IV 1064 for the former, and $C_{los} =0.5$ and the optical depth
profile derived from \ion{P}{5} profile for the latter. The values 
are given in Table 4. 

    The \ion{Si}{2}~$\lambda$1260 and $\lambda$1526 lines are both present in
component D+E.
Despite the factor of $\sim$7 greater optical depth of the shorter wavelength
line, both are shallow.
We interpret this as due to low {\it effective} covering factors for the Si~II
lines resulting from low coverage of the background emission by this absorber.
Figure 10 shows the unocculted flux levels associated with each \ion{Si}{2} line based
on the \ion{P}{5} continuum covering factor and assuming no coverage of
the emission lines. The observed depths are roughly consistent with these.
In Table 4, we also include measurements for the ground-state Fe~II (UV2) and two
metastable states (UV78 and UV263) and an upper limit for the ground-state Fe~III. Although
we expect that there may be strong metastable Fe~III lines, they would lie in a region
of the {\it FUSE} spectrum that is heavily absorbed by Galactic H$_{2}$.

\subsection{Component E$'$}

  Component E$'$ exhibits a rich absorption spectrum, 
including lines from many multiplets and
excited levels, giving diagnostics for detailed $C_{los} - \tau$ analysis
and constraints on the density.
We used the \ion{Si}{2} lines from the ground-state and fine-structure
level ($E_{ll}$=287 cm$^{-1}$) to determine
the covering factor for this component.
The excited level \ion{Si}{2}*~$\lambda$1264.7, 1265.0 lines are strong in E$'$.
Since the shorter wavelength line has nine times greater optical depth, 
it must be saturated with its residual flux due entirely to partial coverage.
This gives the covering factor for both lines directly ($C_{los}\approx$0.8 in the core), 
which we used to measure the column density of the fine-structure level from 
the \ion{Si}{2}*~$\lambda$1265.0 line.
The other \ion{Si}{2} complexes in the STIS spectrum (Si II 1304, Si II* 1309 and
Si II 1526, Si II*1533) have different levels of line emission underlying
their absorption, giving constraints on the individual covering factors
of the continuum and emission line sources.
Joint analysis of these lines reveals $C_c =$ 0.7 -- 1 (continuum)
and $C_l =$ 0.5 -- 0.7 (emission line region).  We used these results to obtain the
ground-state \ion{Si}{2} column density.

  Similar analysis of the \ion{S}{4}*~$\lambda$1073.0, 1073.5 lines gives the 
covering factor and column density for the fine-structure level of \ion{S}{4},
yielding a similar covering factor as the \ion{Si}{2} lines. The 
\ion{S}{4}~$\lambda$1064 ground-state line is consistent with being saturated 
at this covering factor, as required by the large column density of the 
fine-structure level and the maximum ratio of excited to ground state level 
populations (2:1).
Numerous \ion{Fe}{2} lines from the ground-state and four fine-structure levels
in the UV 1, 2, and 3 multiplets are present in component E$'$ in the E230M STIS spectrum.
We derived covering factors and column densities for each level using analysis
of lines with different optical depths.  Our results indicate a somewhat lower 
covering factor ($C_{los} \approx 0.5$ of the continuum source) for this ion than derived from the \ion{Si}{2} and \ion{S}{4} lines.

     For other lines in component E$'$ that do not have independent constraints on the
covering factors, we used the results from the \ion{Si}{2} analysis to derive effective 
covering factors for measurement of column densities (\ion{O}{1}, \ion{Fe}{3}, 
\ion{Ni}{2}, \ion{Mn}{2}, and metastable \ion{Si}{3}). This was done by weighting
the contributions from the continuum and line-emission sources under each line by
their respective covering factors (see Ganguly et al. 1999).  Measured column densities
are listed in Table 5.
Many lines expected to be strong in this component do not exhibit the narrow structure
seen in the above lines (e.g., see the CNO lithium-like doublets in Figure 7a).
We interpret this as due to blending of strong absorption in E$'$ with that of 
the red wing of the broad D+E absorption region.

  Excited level absorption in component E$'$ constrains the density 
in this absorber as well.
The most stringent limit comes from the \ion{Fe}{3} ion,
which has lines detected from the ground-state and three fine-structure levels
between 1122 -- 1132 \AA\ in the {\it FUSE} spectrum.  
The measured column densities for all levels are consistent
with their statistical weight ratios ($g=2J+1$), which occurs 
in the high-density limit.
Based on level population calculations by Keenan et al. (1992),
limits on the column density ratios incorporating measurement 
uncertainties gives a lower limit of $n_e \geq$ 10$^5$~cm$^{-3}$ (note: we
parameterize our photoionization models in terms of $n_{H}$, which is roughly
equivalent to $n_{e}$ over the range in parameter spanned by the models)  .

  Excited-level absorption lines from other ionic species are
consistent with this lower limit.
The \ion{S}{3} J=1 : J=0 column density ratio is consistent with
the high-density limit (2:1 statistical weight ratio); 
the J=2 level is unmeasurable due to contamination of
\ion{S}{3}**~$\lambda$1021.  
The lower limit on the ratio from measurement uncertainties implies 
$n_e \geq$ 10$^{3.8}$~cm$^{-3}$ from the results of Sterling et al. (2005).
The \ion{Si}{2} excited-level/ground-state ratio is also consistent
with the high-density limit (2:1) within uncertainties.

   There are also weak features detected from lines arising from 
highly excited metastable levels of \ion{Si}{3} ($\approx$ 6.2 eV above the ground state).
This is analogous to the metastable structure giving rise to the
\ion{C}{3}*~$\lambda$1175 lines.
We detect weak features associated with the J=2 ($\lambda$1113) and 
J=0 ($\lambda$1108) levels, but only an upper limit on the line associated
with the J=1 level.  
We lack the full set of atomic data to calculate the metastable populations 
as we have for \ion{C}{3}. However, qualitatively these
results indicate the density is not sufficiently high to populate the 
J=1 level. For the \ion{C}{3} ion, the J=1 population exceeds that of J=0
for $n_e \geq$ 10$^{8.5}$~cm$^{-3}$.

\section{Photoionization Models}

The details of the photoionization models used for this study are given in Paper I. To reiterate,
the models were generated using the Beta 5
version of Cloudy (Ferland et al. 1998), which includes
estimated $\Delta$n$=$0 dielectronic recombination (DR) rates for the M-shell states of Fe and
the L-shell states of the third row elements (Kraemer, Ferland, \& Gabel 2004). We assumed
roughly solar elemental abundances (e.g. Grevesse \& Anders 1989) and that
the absorbing gas was free of cosmic dust. The logs of the elemental abundances, relative to H 
by number, are
as follows: He: $-1.00$, C: $-3.47$, N: $-3.92$, O: $-3.17$, Ne: $-3.96$,
Na; $-5.69$, Mg: $-4.48$,  Al: $-5.53$, 
Si: $-4.51$,  P: $-6.43$, S: $-4.82$,  Ar: $-5.40$, 
Ca: $-5.64$,  Fe: $-4.40$, and Ni: $-5.75$.
As per convention, the models are parameterized in
terms of $U$ and $N_{H}$.
We modeled the intrinsic spectral energy distribution as a broken power law of the form $L_{\nu} \propto 
\nu^{\alpha}$ as follows: $\alpha = -1.0$ for energies $<$ 13.6 eV,
$\alpha = -1.3$ over the range 13.6 eV $\leq$ h$\nu$
$<$ 0.5 keV, and $\alpha = -0.5$ above 0.5  
keV. We included a low energy cut-off at $1.24 \times 10^{-3}$ eV (1 mm) and a high energy cutoff
at 100 keV. The luminosity in ionizing photons is  
$Q = 1.1\times10^{53}$ photons s$^{-1}$.

As discussed in Section 1., we initially assumed that there were 5 distinct absorbers. Since the absorbers, except D$+$Ec, 
have high covering factors, it is logical that the more distant
components are irradiated by a continuum that has been filtered by the intervening gas. In Paper I we argued that X-High
was at the smallest radial distance, followed by D$+$Ea, then D$+$Eb, which we assumed to be co-located with D$+$Ec, and,
finally, E$'$, screened by D$+$Eb. Here we have assumed the same order. Since X-High is too highly ionized to produce
detectable UV absorption lines, its only relevance to the UV analysis is that the transmitted spectrum predicted by this
model was used as the input spectrum for modeling component D$+$Ea. Additional components were added if 1) they were required
to fit the UV absorption and 2) their contributions to the X-ray opacity were negligible.

Our approach in modeling the X-ray absorption was, first, to fit the broad-band spectral characteristics 
and, second, to test the details of the models by comparing the predicted and observed absorption lines for
Mg, Si, and S. Similarly, here we compare the predicted ionic column densities to those 
measured from the UV absorption lines. However, we now use the covering factors as constraints in
the determination of the radial distances and densities of the absorber. 

\subsection{Results for Component D$+$E}

\subsubsection{X-ray Derived Model Predictions for Component D$+$E}

In evaluating the model results, it is helpful to review  
the range in velocities and depths (or residual fluxes) for the strongest individual
absorption lines. O~VI, N~V, and C~IV have normalized residual fluxes
of $\sim$ 0.1 from $-$200 to $-$700 km s$^{-1}$, with absorption extending
to velocities $\sim$ $-$1000 km s$^{-1}$ (see Figure 8a). S~VI has a similar residual
flux, but only out to velocities $\sim$ $-$500 km s$^{-1}$ (Figure 8a). C~III appears
to have a velocity range intermediate between the Li-like CNO lines and
S~VI (see Figure 8b). N~III has a profile similar to S~VI, although the residual flux may be
somewhat greater (Figure 8b). Si~IV appears broader than S~VI or N~III, but shows a higher 
residual flux (Figure 8c). Finally S~IV has a velocity profile similar to S~VI, i.e., little
absorption above $-$500 km s$^{-1}$, but has a significantly higher residual
flux (Figures 8d and 10). Note that all of these lines appear to be saturated over much of their
profiles.

Based on this, we would predict that O~VI, N~V, and C~IV would arise primarily in D$+$Ea.
S~VI and C~III would mostly come from D$+$Ea, but their ionic column densities should be 
significantly lower, and we would expect large contributions from D$+$Eb based on the strength
of C~III*. Si~IV would come from D$+$Eb. S~IV could come from a combination of D$+$Eb and D$+$Ec.
The contribution from the former would have to be significantly less than that for Si~IV, 
since the residuals are greater. Sorting out the contributions to N~III is complicated by the
uncertainties in the residual flux: an origin in D$+$Ea, with a column similar to that of C~III, or
an origin in D$+$Eb with a significantly larger column are both possible.    
  
In Table 6, we list the predicted ionic column densities relevant to the UV spectra 
for each of the sub-components of D+E. The model for D$+$Ea predicts O~VI, N~V, and C~IV columns
$>$ 10$^{17}$ cm$^{-2}$. These are large enough that the wings of the lines will begin to become
optically thick, resulting in the broad, square-shaped profiles observed.  Note that
this effect is much stronger in O~VI than in N~V or CIV, as would be expected since O~VI has $\sim$ 10 
times greater optical depth based on the model predictions.
The predicted columns
for S~VI and C~III are $>$ 10$^{15}$ cm$^{-2}$, therefore these lines will be saturated, but with
profiles somewhat narrower than the Li-like CNO lines, in agreement with the data and the suggestion
that the profiles for the stronger lines are the result of extreme saturation. The model
for D$+$Ea predicts smaller columns for N~III, S~IV, and Si~IV. Thus, while there will be some contribution
from D$+$Ea, due to its high covering factor, the absorption is likely dominated by the
other sub-components.

The models for both D$+$Eb and D$+$Ec predict large columns for Si~IV and S~IV. For the former,
the depth of the absorption suggests that the profile is dominated by D$+$Eb. Noting that the
oscillator strengths for S~IV and S~IV* are substantially less than those for the Si~IV doublet, 
the similar predicted column densities would result in weaker/shallower S~IV lines. Hence, we suggest
that the S~IV absorption is dominated by D$+$Ec. The predictions for N~III are consistent with
contributions from each of the three main sub-components, as suggested by the relatively narrow
profile and low residual flux.  

Other than the P~V doublet, there are no absorption features that can be {\it solely} attributed to
sub-component D$+$Ec. Assuming solar phosphorus abundance, the minimum column density required to
produce the observed P~V column is $N_{H}$ $\gtrsim$ 10$^{21.2}$ cm$^{-2}$, while the minimum (screened)
ionization parameter is $\log U \gtrsim -1.2$, consistent with X-ray derived 
model parameters. Although the physical conditions for this sub-component are not well-constrained, the
prediction for S~IV is in agreement with the origin of those lines in D$+$Ec, as noted above. 

None of these models predicts detectable columns for Fe~II, S~II, and Si~II. Since increasing the D$+$Eb column would
result in a prediction of P~V at a much higher covering factor than that observed, the most likely source of these
low ionization species is an additional sub-component, D$+$Ed, which we assumed is screened by D$+$Eb and
parameterized as follows: $N_{H} = 10^{19.5}$ cm$^{-2}$; $\log U = -3.1$. While this model accurately predicts the Si~II and S~II
columns and the upper limit for O~I, it underpredicts Fe~II by a factor of $>$ 5. However, this may be due to uncertainties
in the Fe abundance and/or the Fe~III recombination rates. Furthermore, given the weakness of the Fe lines, 
the measured columns are quite uncertain. However, the main point is that there is a small
amount of low-ionization gas present that was not detectable in the X-ray spectra. 

As we did for component X-High in Paper I, we can determine the maximum radial distance, $R$, for
an absorber by requiring the ratio of the physical depth, $\Delta R = N_{H}/n_{H}$, to $R$ to be less than unity.
Based on our estimate of $Q$ and our model predictions for D$+$Ea, $R = 6.8 \times 10^{20} n^{-1/2}$; from this, 
we find $R \leq$ 5.2 pc. However, it is likely that D$+$Ea is located much closer to the central source, as we will show
in Section 5.1.3.

\subsubsection{Nature of the Partial Line-of-Sight Coverage of the Nuclear Emission}

In Figure 11, we show a deconvolution of the emission-line profiles from the
low-state STIS spectrum obtained in 2000 June 15. In the top panel, we show
He~II $\lambda$1640, which has no absorption and undetectable broad-line
emission. The He~II emission profile clearly shows two components, which we
have fit with Gaussians of width 260 and 1150 km s$^{-1}$ (FWHM), representing
``narrow'' and ``intermediate'' line components. The overall He~II profile is
very similar to that of [O~III] $\lambda$ 5007 from the central
emission-line knot of NGC~4151, observed by STIS with similar aperture size
and spectral resolution (Crenshaw \& Kraemer 2005). In the bottom panel, we
show fits to the C~IV $\lambda\lambda$1548, 1551 emission lines, using the He~II
profiles as templates. We have also added a small contribution from a
spline fit to the broad C~IV emission, which can be detected in the wings.
The fit to the narrow C~IV $\lambda$ 1548 line is excellent, and shows that the
broad D$+$E component does not absorb the narrow-line region, as expected (the
central absorption is from components F and F$'$, which must arise in the host
galaxy or its halo). The fit to the intermediate component of C~IV is also
excellent in the red wing; the lack of emission from the blue wing of this
component indicates that it is absorbed by D$+$E. The observed broad-line and continuum
emission just redward of the intermediate component lies below the model fit due to
Fe~II absorption. Thus, one or more of the
absorption subcomponents in D$+$E must lie outside of the intermediate
emission-line region.

If the intermediate line component is rotationally broadened, we
can estimate its radial location. For a central mass of $\approx$ 10$^{7.66}$ M$_{\odot}$ (Bentz et al.
2006), the radial distance of this component would be $\sim\ 4.6 \times 10^{17}$ cm. However,
there is evidence that much of this emission arises at somewhat smaller radial distances (D.M. Crenshaw \&
S.B. Kraemer, in preparation). The depth of the
absorption in the blue wing of C~IV suggests that it forms in D$+$Ea. If D$+$Ea is roughly co-located
with the emission-line gas, e.g. $\sim$ 0.1 pc, it will have a density of $n_{H}$ $=$ 
10$^{6.8}$ cm$^{-3}$. As noted by K01, the absorbers in the line-of-sight to the central
source are essentially free of dust. Given our derived SED, NGC 4151 has a luminosity above the 
Lyman limit of $\sim$ 10$^{44}$ erg s$^{-1}$. Assuming a dust sublimation temperature of 1500K, the
dust sublimation radius for a source at this luminosity is $\sim$ 0.05 pc (Barvainis 1987); this is
consistent with the Cloudy model predictions for grain survival. 
Given that D$+$Ea may have originated closer to the central source than $\sim$0.1 pc and/or 
the dust sublimation radius was larger when NGC~4151 was in a higher state, the lack of dust in 
this component is understandable. 

Using the model parameters for D$+$Eb from Paper I, and assuming it is at roughly the same radial distance as
D$+$Ea, we obtain a density of $n_{H}$ $=$ 10$^{8.1}$ cm$^{-3}$.
Although $U$ and $N_{H}$ are less well-constrained for D$+$Ec, if we assume that it is co-located with
D+Eb, we obtain a density $n_{H}$ $=$ 10$^{7.4}$ cm$^{-3}$. Interestingly, although we had assumed in Paper I that the C~III* 
absorption arose solely in D$+$Eb, given their physical conditions, each of the sub-components will contribute. 
Since the C~III* lines are sensitive to density effects, we can test our assumptions 
for the distances and densities of the D+E subcomponents by comparing the 
model predictions to the observed C~III* profile, as discussed below.

\subsubsection{C III* $\lambda$1175 Absorption and Density Constraints in Component D+E}

To explore the relative contributions to C~III*, we derived synthetic absorption
profiles based on the photoionization model results for
the low (D$+$Ec), middle (D$+$Eb), and high (D$+$Ea) covering factor regions of D+E.
The predicted C~III* column densities for each sub-component are given in Table 7. 
For contribution from D$+$Ec, we derived an optical depth 
profile for each transition in the \ion{C}{3}*~$\lambda$1175 complex 
by scaling the \ion{P}{5} $\tau$ 
profile (Figure 7a) by the \ion{C}{3}*:\ion{P}{5} $f \lambda N_{ion}$ ratio for each
of the $J$ states.
The optical depth profile of D$+$Ea is more uncertain due to the
heavy saturation of lines associated with this absorber and blending
with the low covering component.
Figure 8 shows several lines have deep, flat-bottomed profiles over these
velocities with weak or negligible absorption outside this velocity
range (e.g., \ion{S}{6}~$\lambda\lambda$933,944 
\ion{N}{3}~$\lambda$989, \ion{N}{3}*~$\lambda$991 and the higher order 
Lyman lines).
The strongest lines (e.g., the CNO lithium-like doublets) extend
to higher outflow velocities ($v \approx -$1000~km~s$^{-1}$),
but are deepest at the lower velocities.
Thus, for the \ion{C}{3}*~$\lambda$1175 model, we assumed a $\tau$ 
profile for each transition that is nearly flat over 
$v \approx-$600 to $-$100~km~s$^{-1}$ and drops steeply
at the boundaries; these were scaled to match the \ion{C}{3}* $J=0,1,2$ model
column densities.
Since the \ion{S}{4}* absorption shows a residual flux intermediate between that determined
from the \ion{P}{5} doublet and that detected in the cores of the CNO Li-like lines, we 
used its absorption profile and the predicted metastable levels to derive 
a model $\tau$ profile for D$+$Eb.

For the combined model absorption, we multiplied the absorption 
equations (Eq.~1) for each subcomponent. For illustrative purposes,
in Figure 12 we show the combined absorption from D$+$Ea and D$+$Ec. The uncertainty in 
how the occulted and unocculted regions of the two subcomponents overlap
as projected against the nuclear emission introduces only a small uncertainty
in these results since the unocculted region of D$+$Ea is small. 
As expected, the absorption from the low and high covering factor regions alone cannot 
reproduce the observed \ion{C}{3}*~$\lambda$1175 absorption strength, based on
the various constraints on the physical conditions and covering factors for
these absorbers. Hence, although the parameterization has changed somewhat, 
this confirms our result from Paper I, specifically, that a contribution from  
another component, specifically D$+$Eb, is required. The combination of the 
three sub-components is shown in Figure 12 (red dashed line). Although the
blue wing absorption is slightly underpredicted, our model matches the overall
absorption profile reasonably well (note that the contribution of E$'$ to the red wing of the profile
has not been included). In fact, these results suggest that the densities may be somewhat higher
than we assumed, hence the components of D$+$E may lie closer to the central source than 0.1 pc. 

\subsection{Results for Component E$^{'}$}

In K01, we were able to successfully model component E$^{'}$ with a single zone, assuming an
unfiltered SED. As noted in section 3, the low-ionization lines detected in 1999 are still
present in 2002. However, the EWs of several lines have increased (see Figure 6), suggesting
a corresponding increase in the column density of this component. 
Also, given the high covering factor that we have derived for D$+$Ea and D$+$Eb,
our previous assumption regarding the ionizing continuum, i.e., that there was no screening by intervening
absorbers, cannot be applicable. Furthermore, the {\it FUSE} spectra reveal the presence of
relatively high-ionization species, including P~V and S~IV, which were not predicted by our
earlier model, and the range of ionization present cannot be replicated
by a single component. Therefore, we modeled the E$^{'}$ absorption using two sub-components,
``a'' and ``b'',
ionized by the transmitted continuum predicted by our model for D$+$Eb, parameterized, respectively,
as follows: $\ log U=-1.74$, $N_{H} = 10^{20.6}$ cm$^{-2}$, and $\log U=-3.64$, $N_{H}= 10^{19}$ cm$^{-2}$.
While we have allowed some flexibility in our parameterization, in order to refine our match to the
UV absorption, the predicted X-ray opacity for E$^{'}$a is similar to what we had 
determined in Paper I. 

The predicted column densities are given in Table 5.
The major discrepancies are the underprediction of Fe~II (see discussion of the model predictions for D$+$E),
S~II, and the overprediction of Mg~II, which arise primarily in E$^{'}$b. Also, the measurements of the ionic column
densities are very sensitive to the assumed covering factor. Another discrepancy is the predicted ratio for the excited
states of O~I. In this case, the measured ratio is unphysical, 
which suggests that our quoted uncertainties are 
underestimated for these weak absorption lines.
The critical test for E$^{'}$a is 
to simultaneously match the measured columns of P~V and C~III$^{*}$ and the limits for S~IV and S~III, since the
zones wherein these ions form overlap. In order to produce the observed C~III$^{*}$ $j=2$ column at the assumed $U$ and N$_{H}$,
we determine the density of E$^{'}$a to be $n_{H}$ $=$ 10$^{6.4}$ cm$^{-3}$. If co-located, E$^{'}$b would have a 
density $n_{H}$ $=$ 10$^{8.3}$ cm$^{-3}$. From the combination of $U$ and
$n_{H}$, we estimate the radial distance of the absorber to be $\approx$ 0.6 pc. 
This is within the loose constraints for E$^{'}$ derived from our analysis of the 1999 spectra (K01). Interestingly,
the relatively small radial distance for E$^{'}$ contrasts with the larger values for the UV absorbers in NGC 3783 (Gabel
et al. 2005), and suggests a connection to the large column of gas associated with component D$+$E.  
    
\section{Comparison with Earlier Epochs}

\subsection{The UV Footprint of the 2000 March X-ray Absorption}

In Paper I, we argued that the greater soft-X-ray opacity observed in the
HETG spectra from 2000 March, compared to 2002 May, resulted from the combined effects of a weaker
ionizing flux and a larger absorbing column. Furthermore, we suggested that
a likely scenario was an increase in the column density of component D$+$Ea. In order to see
if this high column absorber was the present in 1999, we 
generated a model for D$+$Ea assuming 1) the column density derived from the 2000 HETG analyis,
$N_{H} = 10^{22.93}$ cm$^{-2}$, 
and 2) the intrinsic X-ray continuum determined for the 2002 observations (since the UV continua were similar
during 1999 and 2002). The model predicts column densities for 
S~II and Ni~II of 10$^{15.9}$ cm$^{-2}$ and 10$^{15.0}$ cm$^{-2}$, respectively, which are in 
agreement with the measured columns from 1999. The attenuation of the ionizing continuum is so extreme that
the gas in screened components D$+$Eb and D$+$Ec would have recombined, as we suggested in Paper I. 
However, if the column density of D$+$Ea during 2000 March was slightly smaller, e.g. $N_{H} = 10^{22.92}$ cm$^{-2}$,
the S~II and Ni~II column densities would be negligible. In this scenario, these low ionization species would
arise in one of the screened components. In fact, a model for D$+$Eb using the continuum transmitted through this version
of D$+$Ea also predicts the measured column densities for these ions.

The model for D$+$Ea for the 1999 July epoch predicts a Si~IV column of 10$^{17.2}$ cm$^{-2}$, 
hence D$+$Ea would dominate the Si~IV profile (see Table 6). Indeed, the Si~IV profile was broader
in the spectra taken in 1999 July and 2000 March, than in 2002 May, consistent with an extremely 
saturated line. Also, the Si~IV profile is still quite broad 
during 2000 November, at which time the UV continuum was a factor of $\sim$ 1.5 higher than during 1999
and 2002. To test if the broad profile detected in 2000 November was consistent with the same total column
density that was present in 1999, we generated a model for D$+$Ea with N$_{H} = 10^{22.93}$ cm$^{-2}$ (see Paper I) and
an ionizing flux increased by factor of 1.5 over the value used in modeling the 2002 data. The model
predicts N$_{Si^{\rm +3}} = 10^{16.3}$ cm$^{-2}$, which would produce a broad Si~IV profile and is 
consistent with the presence of the large column absorber during the 2000 November epoch. 

However, there is one problem with this scenario. In both the 1999 July and 2000 November spectra, the residual flux in the troughs
of Si~IV over the 
velocities spanned by D$+$Ea was $\sim$ 0.2, similar to that observed in 2002, while the
covering factor of D$+$Ea is $\sim$ 0.9. Although
the column density of D$+$Ea could have been somewhat smaller than we suggested in Paper I, particularly if the
lowest ionization lines observed in 1999 formed in D$+$Eb, weak Si~IV absorption is incompatible with the high soft 
X-ray opacity detected in 2000 March. One possibility is that the extra column present in the earlier data
had a somewhat smaller covering fraction than that of D$+$Ea, hence may not be identical with the sub-component
responsible for the deep Li-like CNO absorption lines.

\subsection{Time Evolution of the UV Absorption}

As discussed above, the column density of the soft-X-ray absorber, which we associated with component
D$+$Ea in Paper I, decreased dramatically between 2000 March and 2002 May. It is plausible that the low-ionization lines detected
in STIS spectra prior to 2002 arise either in D$+$Ea or in one of the associated screened components, most 
likely D$+$Eb. Therefore, the history of this additional absorption can be traced by examining
the changes in low ionization absorption lines, such as Si~II $\lambda$1527 and Si~II* $\lambda$1533, during the
period between 1999 July and 2002 May. As we noted in Section 4, over much of this time the EWs of these
lines were anti-correlated with the UV continuum flux, which indicated that 1) the additional column of gas was present and 2) its
ionization state changed in response to variations in the ionizing flux. There was a dramatic drop in EW between
2001 April and 2002 May, which is best explained by bulk motion of gas out of our line-of-sight, consistent with the
X-ray analysis.

Given the evidence for a very low covering factor for the Si~II lines in 2002 
May (section 4.1), a likely cause for the large decrease in EW is a change in 
covering factor of the background emission since 2001 April. Using the Si~II 
$\lambda$ 1260 line, which is likely saturated, we find that C$_{los}$ was 
$\sim$0.73 in 2001 and $\sim$0.20 in 2002 for the absorber responsible for 
Si~II (likely D+Ed). This suggests that the absorber was in the process of 
moving out of our line of sight to the BLR in 2002. Since the size of the 
absorber is roughly proportional to the square root of the covering factor, 
the absorber moved by about 0.4 times the size of the BLR (6.8 lt days; Metzroth, Onken, \& Peterson [2006]). This results in a 
transverse velocity of v$_T$ $\approx$ 2100 km s$^{-1}$, which is on the same 
order as our lower limit of 2500 km s$^{-1}$ for D+Ea. These velocities
are somewhat higher than the FWHM of the C~IV ILR line, but fall within the velocity range of the
line profile.  This, the transverse velocities are consistent with an origin in the ILR and the
rotational velocities at this distance. However, the connection between the low ionization
absorption present up until 2001 April and the strong X-ray absorption detected in 2000 March, which
results from an additional column of relatively highly ionized gas,
remains unclear.

Regardless of the details of the process, the most recent changes in EW of the UV lines and the drop in opacity of the X-ray
absorption can only be explained by bulk motion of the absorbing gas. This means that
the outflow cannot be purely radial and, indeed, that the dominant component of velocity is transverse to our
line-of-sight. The direction of the transverse velocity
remains unknown, but is linked to the origin and evolution of the outflow. Given that the kinematic studies
suggest that the accretion disk is inclined 45$^{\rm o}$ with respect to our line-of-sight, the transverse
motion of the gas could be directed along a similarly inclined path towards either the accretion disk or 
its rotation axis. The latter would require that the disk were self-gravitating (Paczynski 1978) or, at least, that the local
gravitational potential exceed that of the central point mass, either of which are unlikely (e.g., Yu, Lu, \& Kauffmann 2005). 
The former is
suggestive of the spin-up of magnetic field lines by which jets are accelerated (e.g. Blandford 1983). However, the 
simplest possibility,
which we suggested in Paper I, is that the transverse component is rotational, consistent with an origin in a disk-driven wind.
In this case, the tranverse velocity would be on the same order as the rotational velocity at the launch pad of the flow. 
However,  possible problem with this scenario is that it requires D$+$Ea to originate at a much smaller radial
distance ($\sim$ 10$^{16}$ cm) than that of the putative ILR, although the ILR C~IV is clearly covered by this
absorber.

\section{Summary} 

Based on our modeling of the UV and X-ray absorbers, we have developed a set of constraints on the physical
conditions and evolution of the absorbers. The main results are as follows.

1) The bulk of the soft X-ray absorption and the strong Li-like CNO absorption lines 
associated with kinematic component D$+$E can be modeled as forming in a single component, D$+$Ea. 
There are three additional lower-ionization sub-components of D$+$E, which do not contribute significantly to the X-ray absorption. 

2) Our deconvolution of the C~IV profile indicates that at least some of the absorbing gas associated with D$+$E must
lie outside the ILR. If the ILR emission arises at a radial distance of 0.1 pc, from our derived values for $U$ we can constrain
the densities of the absorbers. Using these densities, and the predicted ionic columns from the metastable states of C~III when these
densities are input to the models,
we were able to fit the observed C~III* profile, hence our assumptions are self-consistent. 

3) The EWs of the low ionization lines associated with D$+$E have varied over the period from 1999 July to 2002 May. Before
2001 April, these changes were correlated with variations in the UV continuum, as expected for changes in ionic 
columns due to the response of the gas to the ionizing continuum. However, between 2001 April and 2002 May, the drop
in the EWs of these lines are more suggestive of bulk motion, as does the change in the X-ray opacity between between 2000 March
and 2002 May (Paper I). The weakness of the metastable Fe~II lines in 2002, compared to 1999 when NGC 4151 was in a similar
flux state, must also be due to bulk motion. However, it is uncertain how the column densities of the individual sub-components of D$+$E have varied.

4) We have obtained constraints on the transverse velocity, $v_{T}$ $\approx$ 2100 km s$^{-1}$, for the low ionization absorption 
associated with D$+$E, which we dubbed D$+$Ed. While this is roughly comparable to the lower limit derived for D$+$Ea in Paper I, the
connection between these sub-components and the origin of the strong low ionization absorption present in the 1999 July STIS spectra
is not clear.

In summary, while we have modified some of the details of the models used in Paper I, and included additional components of low-ionization gas, 
the UV analysis has confirmed our two principal conclusions. First, the bulk of the absorption in NGC 4151 is much closer to the
central source than in other Seyferts, e.g. NGC 3783 (Gabel et al. 2005b) and, second, that the variations in the absorption are due to
both ionization and bulk motion. The latter could be a consequence of an origin within a rotating disk, which is consistent with our
suggestion that the outflow is MHD-driven. 

An interesting characteristic of the absorbers which is yet to be addressed is the origin of low covering factor sub-components,
such as D$+E$c. Although one
can envision denser, lower ionization knots embedded in a more tenuous flow, the dynamics of such a system are not straightforward. One major problem
is that there are multiple components, characterized by different $U$'s and, thus different Force Multipliers (see Paper I and references therein),
with the same radial velocities. While this may be consistent with an MHD flow, it requires that the
low ionization knots are coupled to the flow in such a way that they are not radiatively driven through the more
tenuous gas.

\acknowledgments

Support for proposal GO-9272 was provided by NASA through a grant from the 
Space Telescope Science Institute, which is operated by the Association of 
Universities for Research in Astronomy, Inc., under NASA contract NAS
5-26555. Some of the data presented in this paper were obtained from the
Multimission Archive at the Space Telescope Science Institute (MAST). This
research has made use of NASA's Astrophysics Data System. We thank Marianne Vestergaard
and Fred Bruhweiler for useful discussions.

\clearpage

\begin{deluxetable}{lrr}
\tablecolumns{3}
\footnotesize
\tablecaption{X-ray Absorber Model Parameters}
\tablewidth{0pt}
\tablehead{
\colhead{Component} & \colhead{$\log U^{a}$} & \colhead{$\log N_{H}$}}
\startdata
X-High & 1.05 & 22.5 \\
D$+$E{\it a} & $-$0.27 & 22.46 \\
D$+$E{\it b} & $-$1.67 & 20.8  \\
D$+$E{\it c} & $-$1.08 & 21.6  \\
E$^{'}$ & $-$1.59 & 20.8  \\
\\
\enddata
\tablenotetext{a}{The ionization parameters for the screened components include the effects
of absorption by intervening gas (see Paper I for a detailed description).}
\end{deluxetable}

\begin{deluxetable}{cccrl}
\tablecolumns{5}
\footnotesize
\tablecaption{{\it HST} STIS Echelle Observations of NGC 4151}
\tablewidth{0pt}
\tablehead{
\colhead{Grating} & \colhead{Coverage} &
\colhead{Resolution} & \colhead{Exposure} & \colhead{Date} \\
\colhead{} & \colhead{(\AA)} &
\colhead{($\lambda$/$\Delta\lambda$)} & \colhead{(sec)} & \colhead{(UT)}
}
\startdata
E140M &1150 -- 1710 &46,000 &5546 &1999 July 19 \\
E230M &1615 -- 2360 &30,000 &2362 &1999 July 19  \\
E230M &2275 -- 3100 &30,000 &2304 &1999 July 19 \\
E140M &1150 -- 1710 &46,000 &9445 &2000 March 3 \\
E230M &2275 -- 3100 &30,000 &2560 &2000 March 3 \\
E140M &1150 -- 1710 &46,000 &2200 &2000 June 15 \\
E140M &1150 -- 1710 &46,000 &2200 &2000 November 5 \\
E140M &1150 -- 1710 &46,000 &2242 &2001 April 14 \\
E140M &1150 -- 1710 &46,000 &7642 &2002 May 8 \\
E230M &1615 -- 2360 &30,000 &2700 &2002 May 8 \\
E230M &2275 -- 3100 &30,000 &2700 &2002 May 8 \\
\enddata
\end{deluxetable}

\begin{deluxetable}{ccrl}
\tablecolumns{4}
\footnotesize
\tablecaption{{\it FUSE} Observations of NGC 4151}
\tablewidth{0pt}
\tablehead{
\colhead{Coverage} &
\colhead{Resolution} & \colhead{Exposure} & \colhead{Date} \\
\colhead{(\AA)} & \colhead{($\lambda$/$\Delta\lambda$)} & \colhead{(sec)}
& \colhead{(UT)}
}
\startdata
905 -- 1180 &20,000 &20,596 &2000 March 5 \\
905 -- 1180 &20,000 &13,612 &2001 April 8 \\
905 -- 1180 &20,000 &48,892 &2002 May 28 \\
905 -- 1180 &20,000 &5,953 &2002 June 1 \\
\enddata
\end{deluxetable}

\clearpage
\begin{deluxetable}{lllll}
\tablewidth{0pt}
\tablecaption{Measured Ionic Column Densities\tablenotemark{a} for Component D+E}
\tablehead{
\colhead{Ion} & \colhead{$E_{ll}$ (cm$^{-1}$)} & \colhead{$C_{los}=0.5$} &\colhead{$C_{los}=0.8$} 
&  \colhead{$C_{los} =1.0$}}
\startdata
H~I (Ly-7) & & & & $\geq 7.1e02$\\
C~II & & &  $\geq 6.1$ & \\
C~III$^{*}$\tablenotemark{b} $J=0$ & 52367 & & $<$ 3.2 & \\
~~~~~~~~ $J=1$ & 52390 &  & $<$ 2.0  & \\
~~~~~~~~ $J=2$ & 52447 & & $\geq$ 6.3  & \\
O I & & $\leq$1.5 &  & \\
Mg~II (2803) & & $\geq 0.4$ & &  \\
Al~III & & & $\gtrsim$ 0.7 & \\
Si~II (1526) & & 1.1 & & \\
Si~II (1533) & & 2.0 & & \\ 
Si~II total  & & 3.1 & & \\ 
Si~III (1113)&  & & $\geq 1.2$ & \\
Si~IV (1403)&  & & & $\geq 3.6$\\
P V & & 3.5 &  & \\
S II (1253)&  & $\leq$0.8 &  & \\
S III $J=0$ & 0  & $\leq$ 4.5  & $\leq 3.0$ &\\
S III* $J=1$ & 299 & 11 & 8.0 &\\
S III** $J=2$ & 833 & 13\tablenotemark{c} & 9.0 & \\
S III total & &  $\leq$ 29 & $\leq 20$ &\\
S~IV & & & $>$15 &  \\
S~IV$^{*}$ & 951 & & $>$29 & \\
S~IV total &   & $>$44 &   \\
Fe~II (2383; UV2) & & 0.6 & & \\
Fe~II (2664; UV263)&  & 0.4 & & \\
Fe~II (2984; UV78) & & 0.9 & & \\
Fe~III (1123) &  & $\leq 1.5$ & & \\  
\enddata
\tablenotetext{a}{Column densities and upper limits were measured using 
covering factors of 0.5, 0.8, and 1.0. For the values determined using $C_{los}=0.5$, we assumed a profile derived from the \ion{P}{5} doublet.
We have not listed specific measurement uncertainties, since the primary source of
error is the uncertainty in the covering factors. Values are in units of 10$^{14}$~cm$^{-2}$. Column densities refer to the ground state ions, unless otherwise noted.}
\tablenotetext{b}{The C~III$^{*}$ column densities for $C_{los}=0.8$ were determined, using the S~IV$^{*}$ profile, assuming
{\it all} of the absorption arises in this sub-component (see text).}
\tablenotetext{c}{Potential contamination with Ly$\beta$ components A,B}
\end{deluxetable}
\clearpage

\clearpage
\begin{deluxetable}{llllll}
\tablewidth{0pt}
\tablecaption{Measured and Predicted Ionic Column Densities\tablenotemark{a} for Component E$'$}
\tablehead{
\colhead{Ion} & \colhead{$E_{ll}$ (cm$^{-1}$)} & \colhead{Measured Column Density}
& \colhead{E$'$a} & \colhead{E$'$b} & \colhead {Predicted Total} } 
\startdata
C II & & $\geq$ 1.2 & 6.68 & 17.4 & 24.1 \\
C III* $J=2$ & & 0.70 $\pm$ 0.20 & 0.85 & 0.01  & 0.86 \\
O I & 0 & 0.84 $\pm$ 0.30 & $-$\tablenotemark{b} & 1.05 & 1.05\\
O I & 158 & 0.33 $\pm$ 0.10  & $-$ & 0.62 & 0.62 \\
O I & 226 & 0.52 $\pm$ 0.15 & $-$ & 0.21 & 0.21 \\
Mg II & & 1.62 $\pm$ 0.17 & 1.79 & 1.97 & 3.76\\
Al II & &  0.22 $\pm$ 0.02 & 0.07 & 0.24 & 0.31\\
Al III & &  $\geq$ 0.11 & 0.92 & $-$ & 0.92\\
Si II & 0 & 1.2 $\pm$ 0.3& 0.17 & 0.62 & 0.79\\
Si II* & 287 & 1.9 $\pm$ 0.5& 0.33 & 1.20 & 1.53\\
Si III* $J=0$ & 52759 & 0.03 $\pm$ 0.01 & & & \\
Si III* $J=2$ & 52840 & 0.04 $\pm$ 0.01 & & & \\
Si III total &  & $>$ 0.07 & 24.4 & 1.27 & 25.7 \\
P V &   & 0.11 $\pm$ 0.05 & 0.11 & $-$ & 0.11 \\
S II &   & 2.9 $\pm$ 0.7 & 0.12 & 0.70 & 0.82 \\
S III $J=0$ & 0 & 2.1 $\pm$ 0.5 & & &  \\
S III* $J=1$ & 299 & 5.5 $\pm$ 1.5  & & & \\
S III total &  & $\leq$ 17\tablenotemark{c} & 13.1 & 0.70 & 13.8 \\
S IV & 0 & $\geq$5.0 &  & & \\
S IV*\tablenotemark{d} & 951 & 19 $\pm$ 5 & & & \\
S IV total &  & $\geq$ 24.0 & 35.6 & $-$ & 35.6 \\
Fe II & 0 & 1.6 $\pm$ 0.5 & & & \\
Fe II* & 385 & 0.57 $\pm$ 0.15 & & & \\
Fe II** & 668 & 1.1 $\pm$ 0.28 & & & \\
Fe II*** & 863 & 0.81 $\pm$ 0.22 & & & \\
Fe II**** & 977 & 0.41 $\pm$ 0.13 & & & \\
Fe II total &  & $\geq$ 4.49 & $-$ & 0.75 & 0.75 \\
Fe III & 0 & 1.8 $\pm$ 0.45 & & & \\
Fe III* & 436 & 1.3 $\pm$ 0.35 & & & \\
Fe III** & 739 & 0.86 $\pm$ 0.25 & & & \\
Fe III*** & 932 & 0.51 $\pm$ 0.15 & & &  \\
Fe III total &  & $\geq$ 4.47 & 6.55 & 2.90 & 9.45\\
Mn II\tablenotemark{e} &   & 0.04 $\pm$ 0.02 & & &  \\
Ni II &   & 0.18 $\pm$ 0.06 & $-$ & 0.11 & 0.11\\
\enddata
\tablenotetext{a}{All column densities were measured using the covering
factors derived from the \ion{Si}{2} lines; values are in units of 10$^{14}$~cm$^{-2}$.
Column densities refer to the ground state ions, unless otherwise noted.}
\tablenotetext{b}{Dash denotes predicted column densities $<$ 10$^{12}$~cm$^{-2}$.}
\tablenotetext{c}{Sum of measured J=0 and J=1 level column densities and upper limit 
on J=2 level based on the statistical weight ratio in the high density limit.}
\tablenotetext{d}{The individual S~IV* lines cannot be resolved in this component.}
\tablenotetext{e}{Manganese was not included in the photoionization models.}
\end{deluxetable}
\clearpage

\clearpage
\begin{deluxetable}{lllll}
\tablewidth{0pt}
\tablecaption{Predicted Ionic Column Densities\tablenotemark{a} for Component D+E}
\tablehead{
\colhead{Ion} & \colhead{D$+$Ea} &\colhead{D$+$Eb} 
&  \colhead{D$+$Ec} &\colhead{D$+$Ed}}
\startdata
H~I & 1.06e03 & 1.38e03 & 2.51e03 & 2.58e03\\
C~II & 0.01 & 7.71 & 5.58 & 25.1 \\
C~III & 44.6 & 8.01e02 & 2.15e03 & 80.5 \\
C~IV & 2.22e03 & 1.31e03 & 1.11e04 & 1.73\\
N~III & 8.32 & 2.84e02 & 7.55e02 & 28.2\\
N~V & 1.55e03 & 3.44 & 41.5 & $-$ \\
O~I  & $-$ & $-$ & $-$ & 0.66 \\
O~VI & 3.08e04 & 0.45 & 21.0 & $-$ \\
Mg~II & $-$ & 2.16 & 3.83 & 3.20\\
Al~III & $-$ & 1.27 & 2.15 & 0.31 \\
Si~II & $-$  & 0.52 & 0.15 & 2.26 \\
Si~III & $-$ & 31.3 & 25.2 & 7.06\\
Si~IV & 0.52 & 75.7 & 2.30e02 & 0.47\\
P~V & 0.63 & 0.24 & 3.50 & $-$ \\
S~II & $-$ & 0.15 & 0.05 & 0.97 \\
S~III & 0.02 & 17.9 & 22.6 & 3.59 \\
S~IV & 2.05 & 54.1 & 2.59e02 & 0.18 \\
S~VI & 72.4 & 2.86 & 26.9 & $-$ \\
Fe~II & $-$ & $-$ & $-$ & 0.37 \\
Fe~III & $-$ & 3.62 & 5.53 & 6.50 \\
Ni~II & $-$ & $-$ & $-$ & 0.07 \\ 
\enddata
\tablenotetext{a}{Values are in units of 10$^{14}$~cm$^{-2}$.}
\
\end{deluxetable}
\clearpage

\clearpage
\begin{deluxetable}{llll}
\tablewidth{0pt}
\tablecaption{Predicted C~III* Column Densities\tablenotemark{a} for Component D+E}
\tablehead{
\colhead{C~III* state} & \colhead{D$+$Ea} &\colhead{D$+$Eb} 
&  \colhead{D$+$Ec}}
\startdata

$J=0$ & 0.30 & 0.50 & 1.20 \\
$J=1$ & $<$ 0.01  & 0.25 & 0.18 \\
$J=2$ & 1.89 & 3.09 & 7.64 \\

\enddata
\tablenotetext{a}{Values are in units of 10$^{14}$~cm$^{-2}$.}

\end{deluxetable}

\clearpage

\figcaption[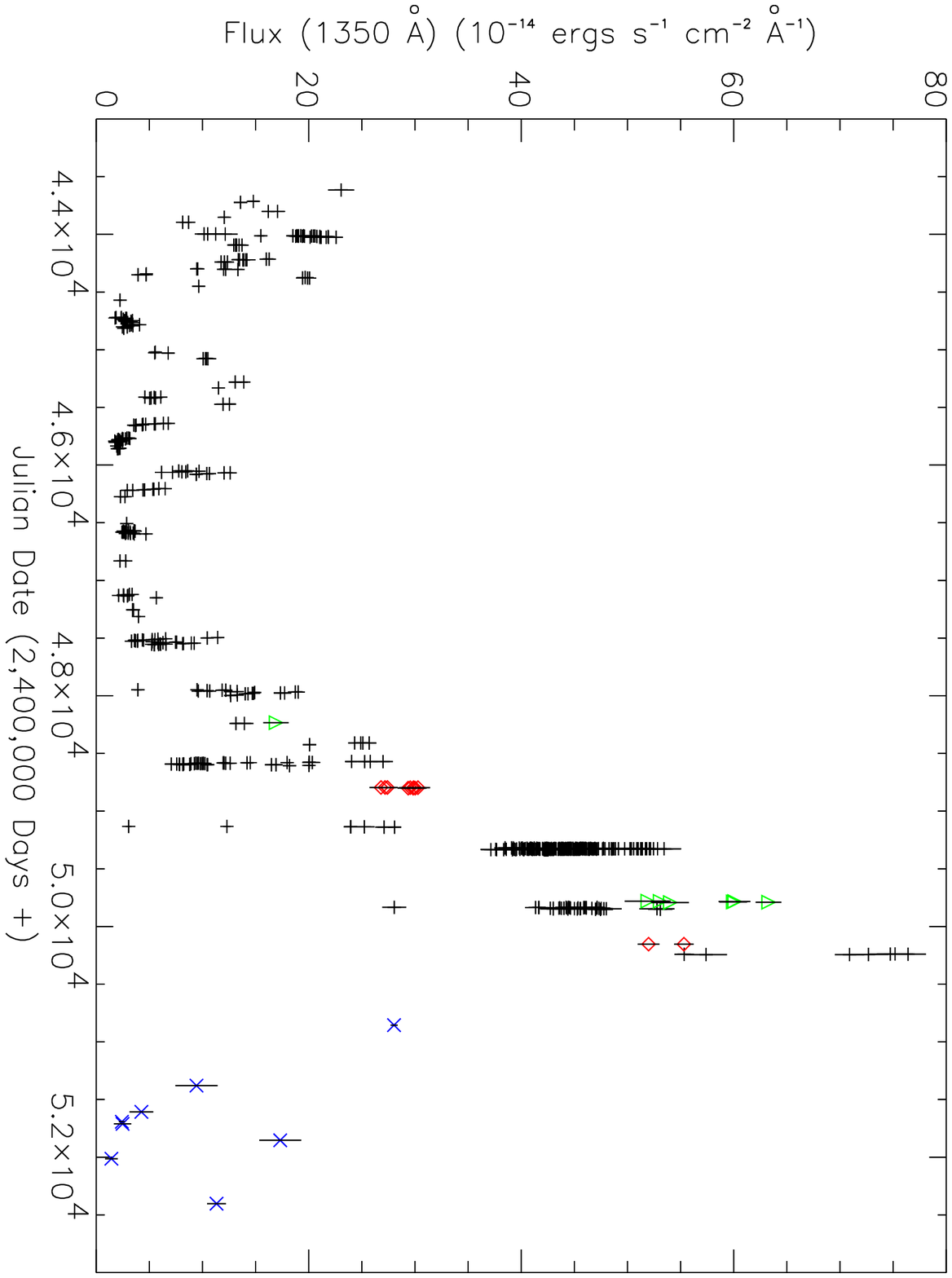]{Far-UV continuum light curve of NGC~4151. Fluxes 
(10$^{-14}$ ergs s$^{-1}$ cm$^{-2}$ \AA$^{-1}$) at 1350~\AA\  are plotted as a
function of Julian date. The symbols are as follows: black pluses -- {\it IUE},
green triangles -- {\it FOS}, red diamonds -- {\it HUT}, blue X's -- STIS. The
first STIS point is a low-dispersion observation, and the fourth point is
actually a superposition of two observations (one low dispersion and one
echelle) separated by only 18 days. The remaining STIS points are echelle
observations. Vertical lines indicate the error bars ($\pm$ one sigma).}

\figcaption[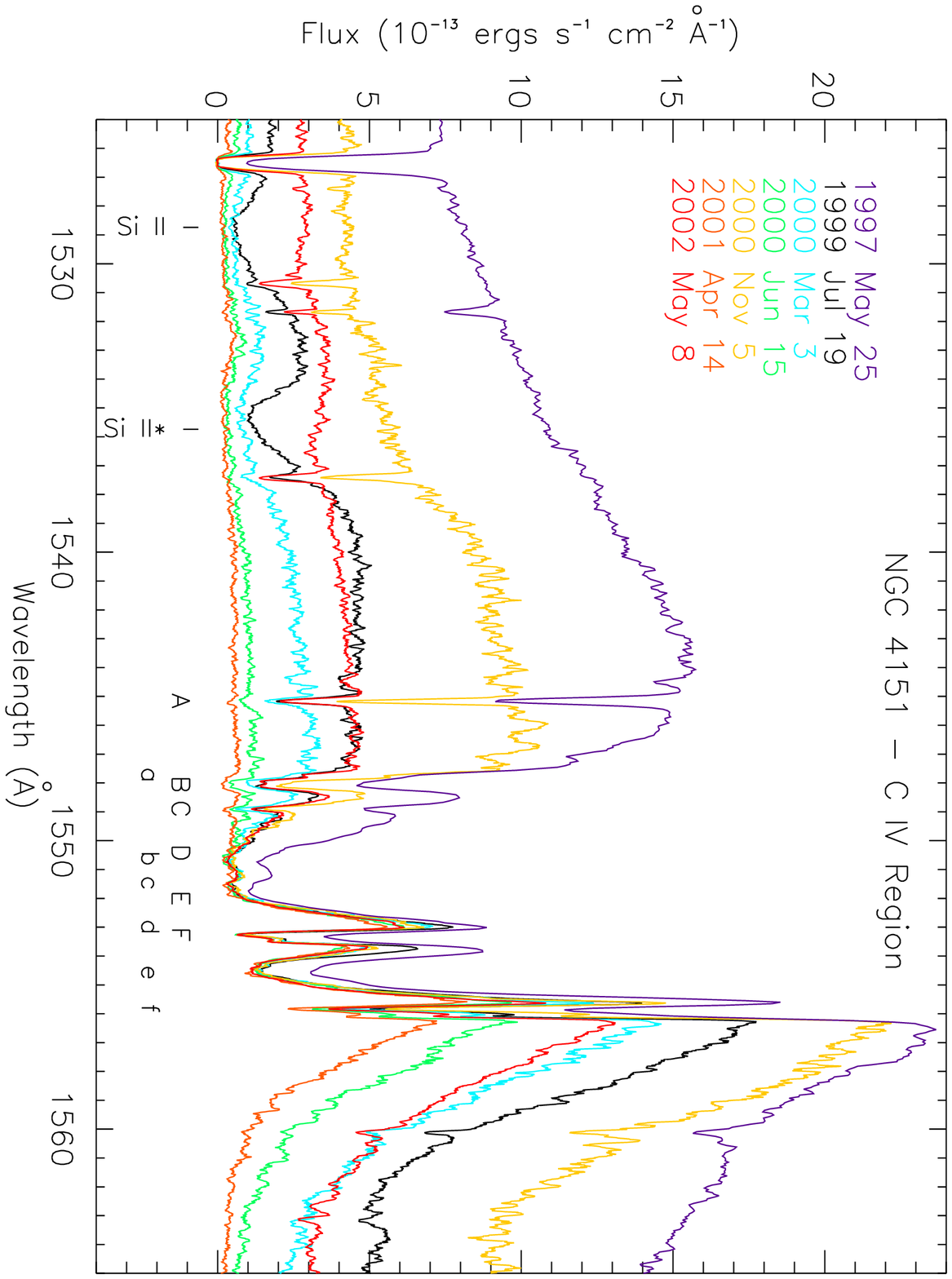]{STIS echelle spectra of NGC~4151 in the C~IV region and a
STIS G140M spectrum of the nucleus obtained in 1997. Approximate locations of
the kinematic components defined by
Weymann et al. (1997) are labeled in capital letters for C IV $\lambda$1548
and small letters for C~IV $\lambda$1551. The D$+$E components in Si~II
$\lambda$ 1527 and Si~II* $\lambda$1533 are also labeled. Note the dramatic
decrease in the Si~II equivalent widths between the 1999 and 2002 STIS echelle
observations, despite the similar continuum levels (see Figure 1).}

\figcaption[f3.eps]{Equivalent widths of three absorption lines associated with
kinematic component A. The continuum light curve from the echelle observations
is given in the bottom panel. Missing points on particular dates indicate nondetection
of the line, due to the extremely low flux.}

\figcaption[f4.eps]{Same as in Figure 1 for component C.}

\figcaption[f5.eps]{Same as in Figure 1 for component D$+$E.}

\figcaption[f6.eps]{Same as in Figure 1 for component E$'$.}

\figcaption[civ_dprime.eps]{STIS echelle spectra of NGC~4151, from 1999 July (black) and 2002 May (green), in the C~IV region.
The spectra are plotted as observed flux versus velocity and show the approximate location of component D$^{'}$.}

\figcaption[fig_apj_profsa.ps]{
   Normalized intrinsic absorption velocity profiles of some
key lines in the 2002 May {\it HST}/STIS and {\it FUSE} 
spectra of NGC 4151.  Centroid velocities of different kinematic 
components identified in earlier studies are marked with dashed 
vertical lines. A model of the Galactic H$_2$ lines present in 
the {\it FUSE} spectrum is plotted (dotted features). Additional
contaminating absorption features are identified with tickmarks.}

\figcaption[fig_apj_p5chisqc.ps]{
 Solution to covering factor and optical depth profiles for the
\ion{P}{5}~$\lambda\lambda$1118,1128 doublet in component D+E.  
$C_{los}- \tau$ solutions and 1$\sigma$ uncertainties were derived
from $\chi^2$ analysis, as described in the text. Contaminated 
regions of the \ion{P}{5}~$\lambda$1118 feature are plotted
with a dotted line.  The solution reveals low covering factor 
($C_{los} \approx$0.5) consistently over the independent solutions
in each velocity bin.}

\figcaption[fig_apj_delc_1-C.ps]{
Normalized absorption profiles for lines in component D+E
compared with the unocculted flux levels predicted for the low covering factor region.
Unocculted flux levels shown (red) were derived based on
the covering factor of the continuum source from the \ion{P}{5} 
doublet and for no coverage of the line emission.  
Contaminating absorption features are marked with tick marks.}

\figcaption[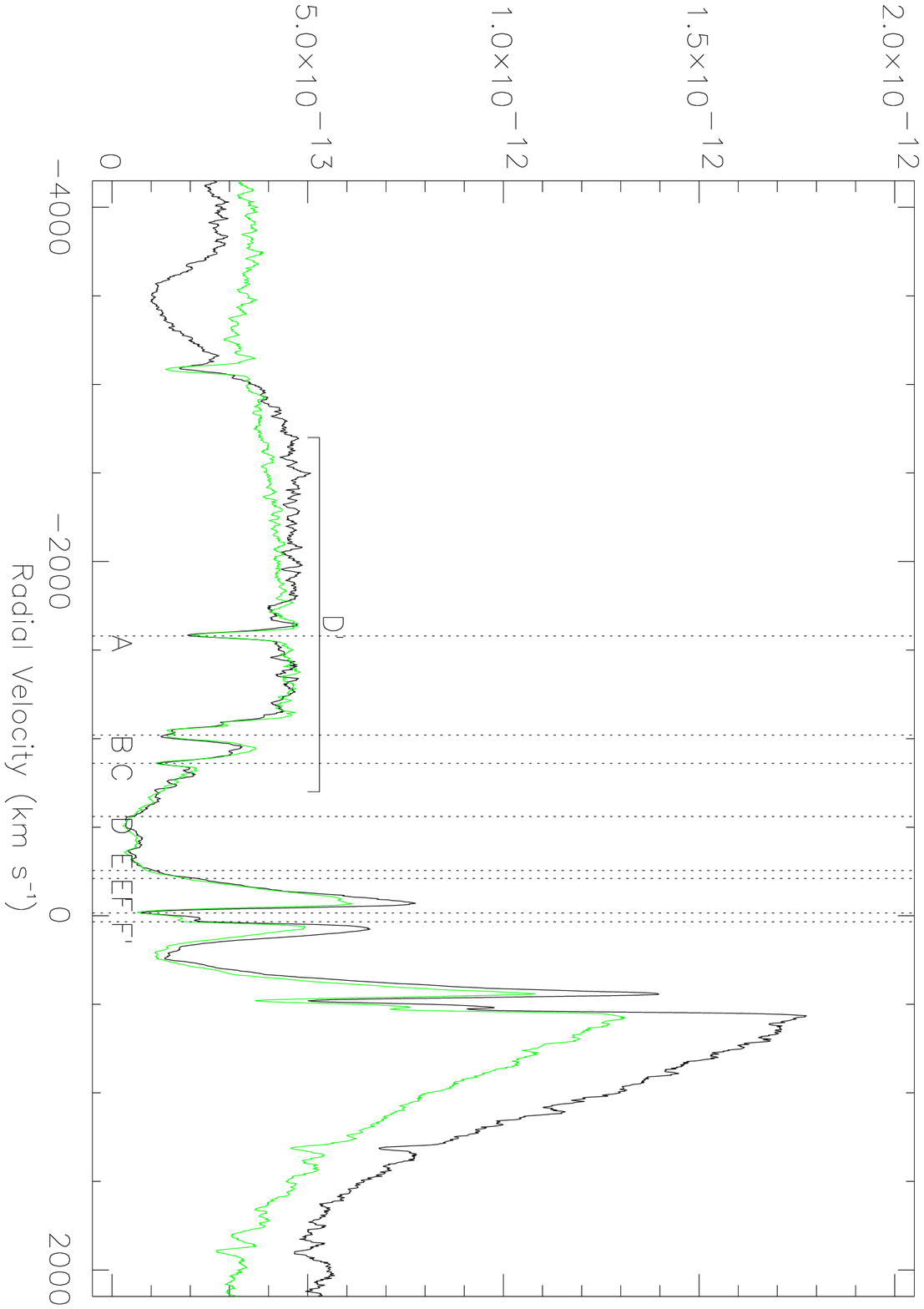]{Top: deconvolution of the narrow and intermediate
components of He~II $\lambda$1640 emission in a low-state spectrum (2000 June
15). The blue dotted, green dashed, and black solid lines show the narrow,
intermediate, and total emission profiles plus the continuum fit, respectively.
Bottom: fits to the narrow and intermediate components of C~IV emission using
the He~II components as templates, as well as a spline fit to the broad
component. The blue dotted, green dashed, red dotted-dashed, and black solid
lines show the narrow, intermediate plus broad, broad, and total emission
profiles plus continuum fit, respectively.}

\figcaption[fig_apj_c3ms_3cb.ps]{
 The \ion{C}{3}*~$\lambda$1175 absorption in component D+E.  
A model of the combined
absorption for the low (D+E$_{c}$) and high (D+E$_{a}$) covering
factor subcomponents based on density constraints and photoionization
model results (blue dashed line) underestimates the observed 
\ion{C}{3}*~$\lambda$1175 absorption (black).
A representative model including absorption from a third component (D+E$_{b}$) that
meets other constraints is shown (red dashed line).
The location of the six lines in the \ion{C}{3}~$\lambda$1175 complex for
component E' are identified with tickmarks and labeled by their metastable level ($J$), 
showing the mismatch in the red-wing of D+E is due to absorption from this
component.}

\clearpage
\begin{figure}
\plotone{f1.eps}
\\Fig.~1.
\end{figure}

\clearpage
\begin{figure}
\plotone{f2.eps}
\\Fig.~2.
\end{figure}

\clearpage
\begin{figure}
\epsscale{.9}
\plotone{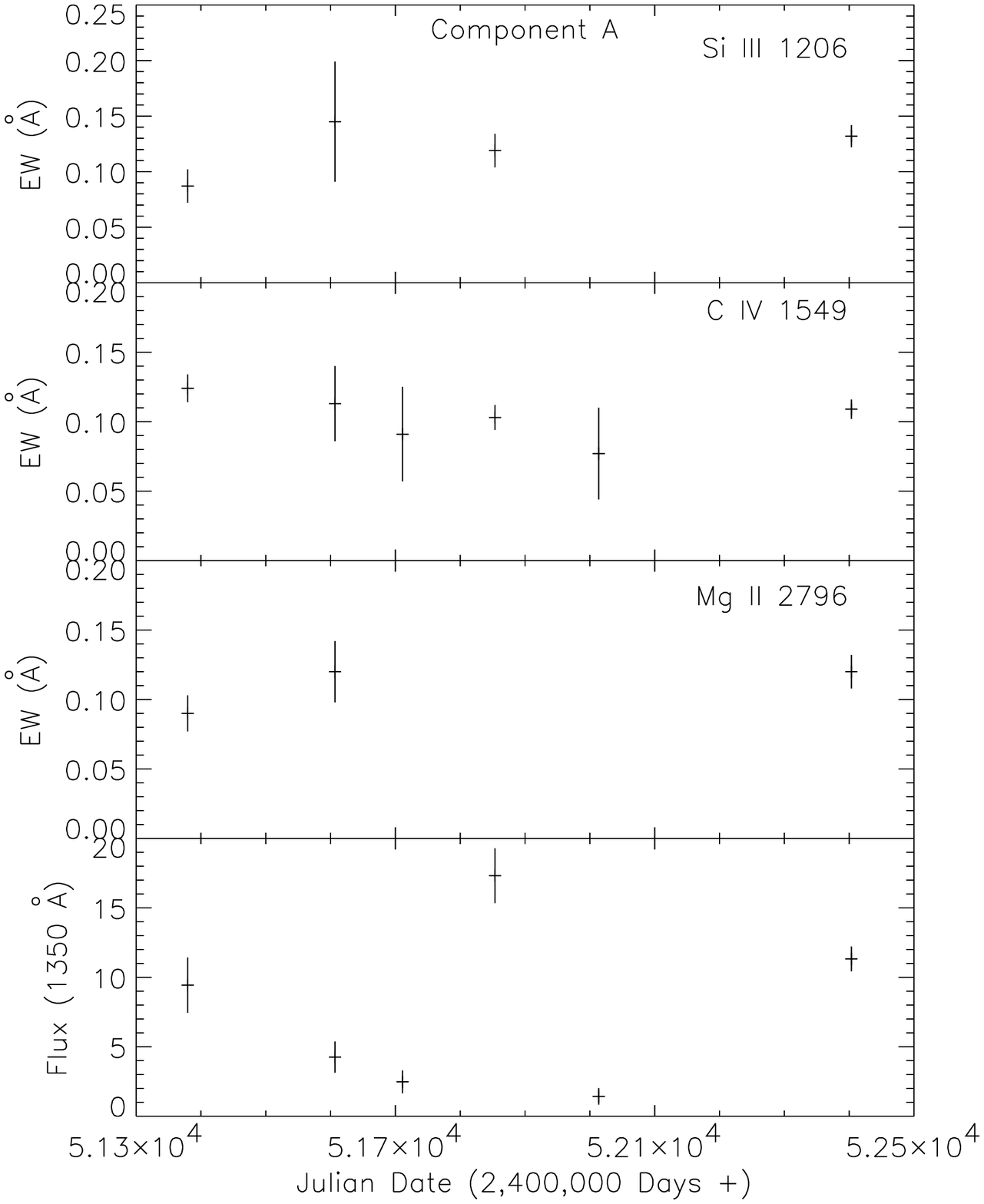}
\\Fig.~3.
\end{figure}

\clearpage
\begin{figure}
\plotone{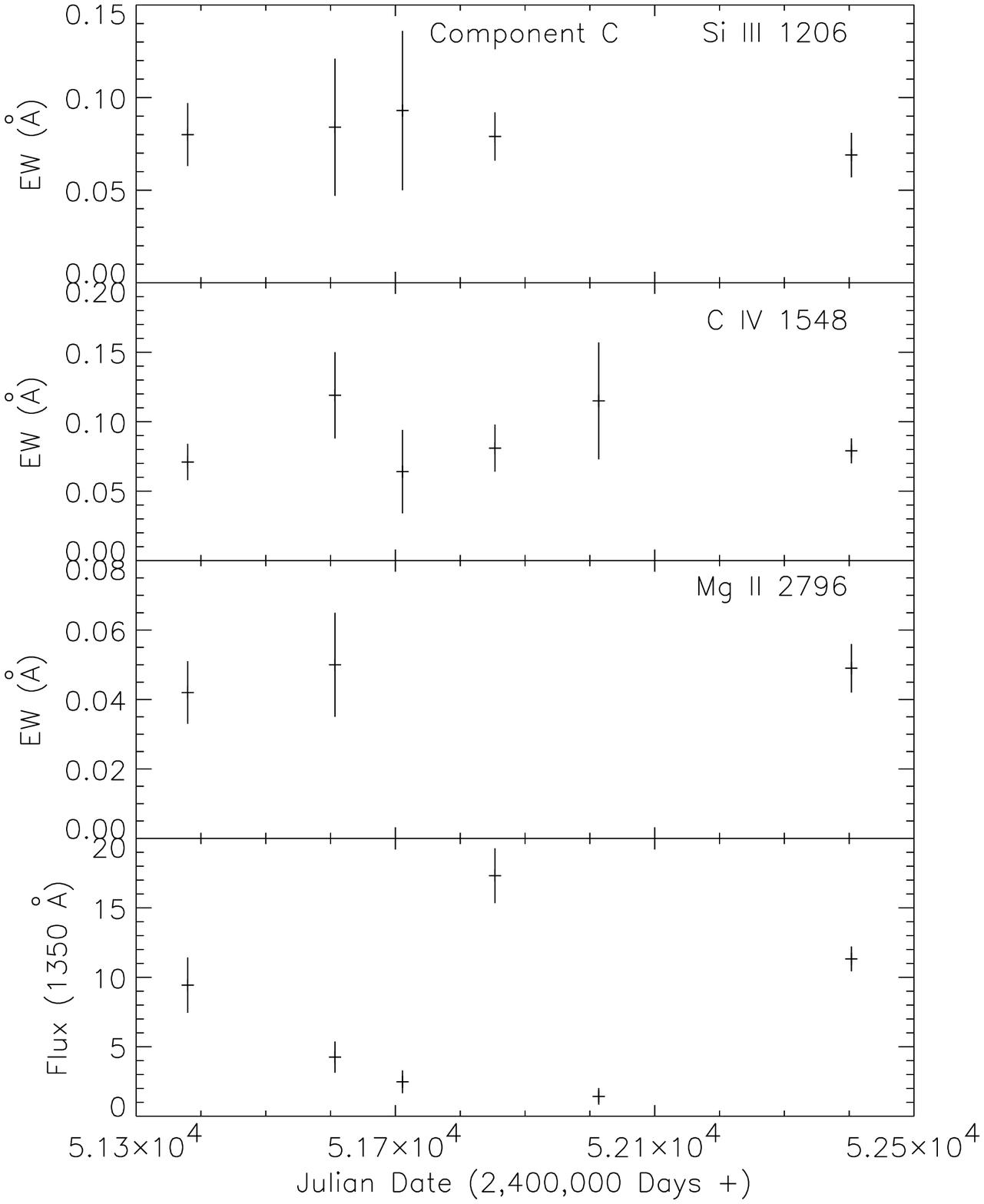}
\\Fig.~4.
\end{figure}

\clearpage
\begin{figure}
\plotone{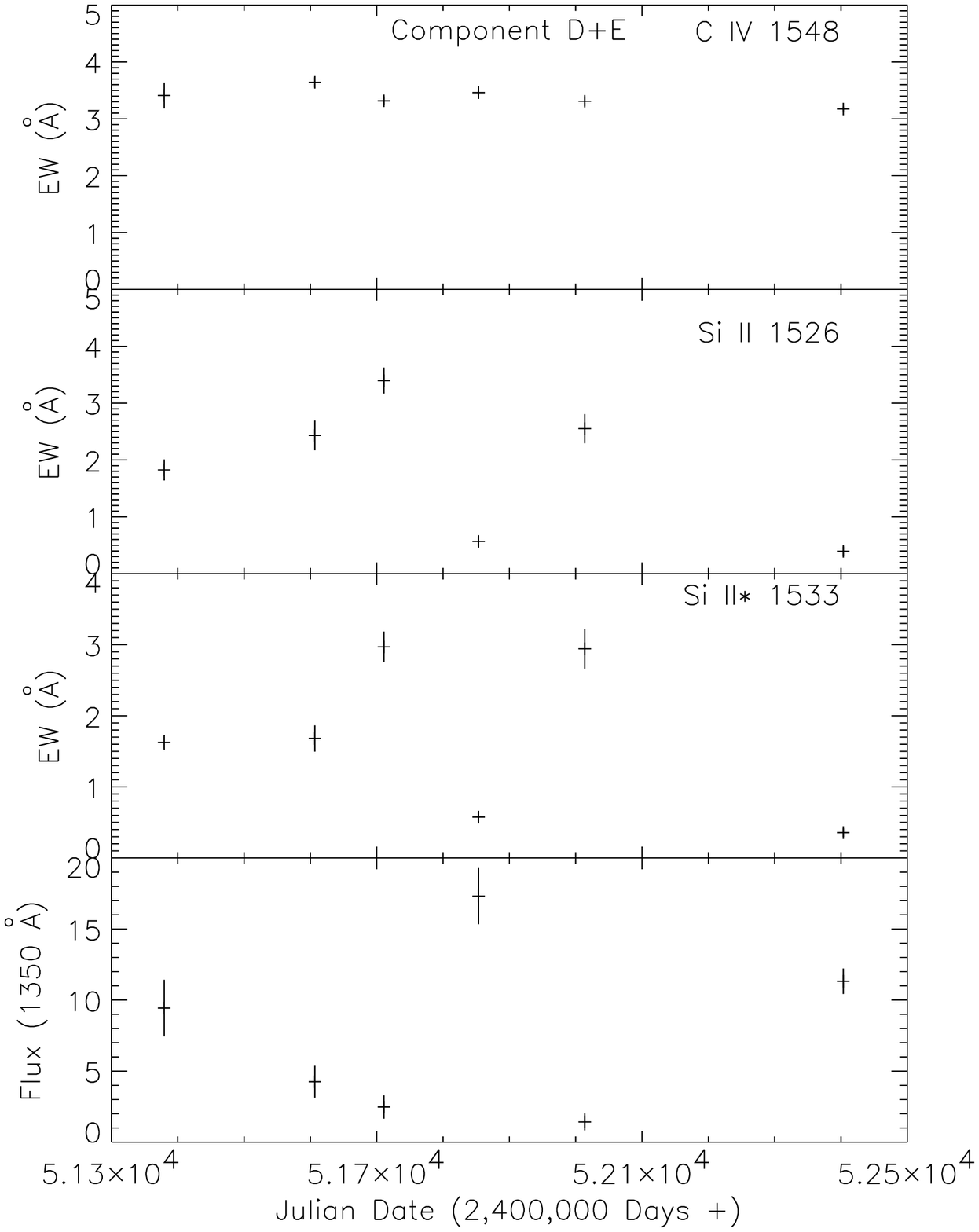}
\\Fig.~5.
\end{figure}

\clearpage
\begin{figure}
\plotone{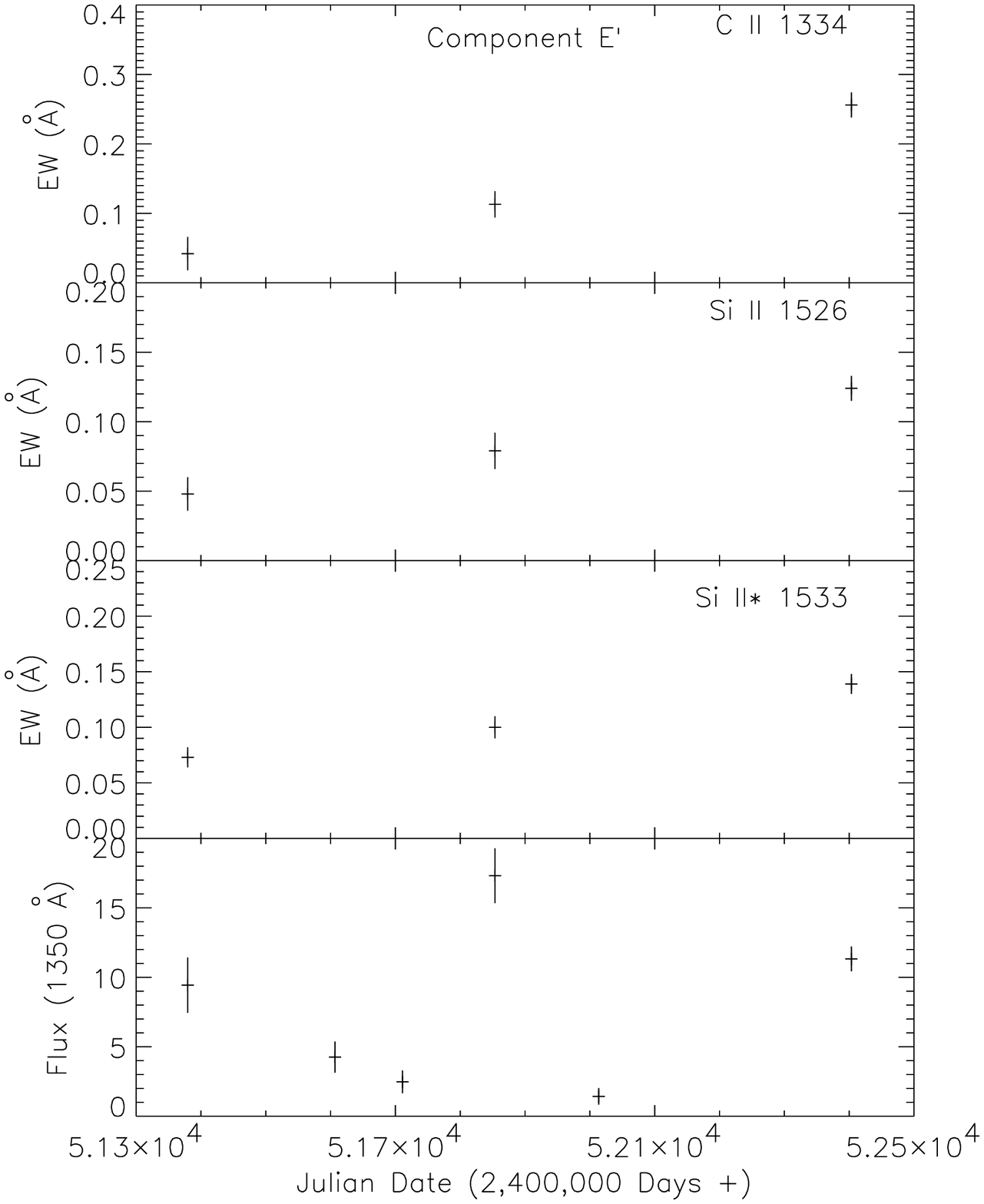}
\\Fig.~6.
\end{figure}

\clearpage
\begin{figure}
\plotone{f7.eps}
\\Fig.~7.
\end{figure}

\begin{figure}
\epsscale{0.6}
\plotone{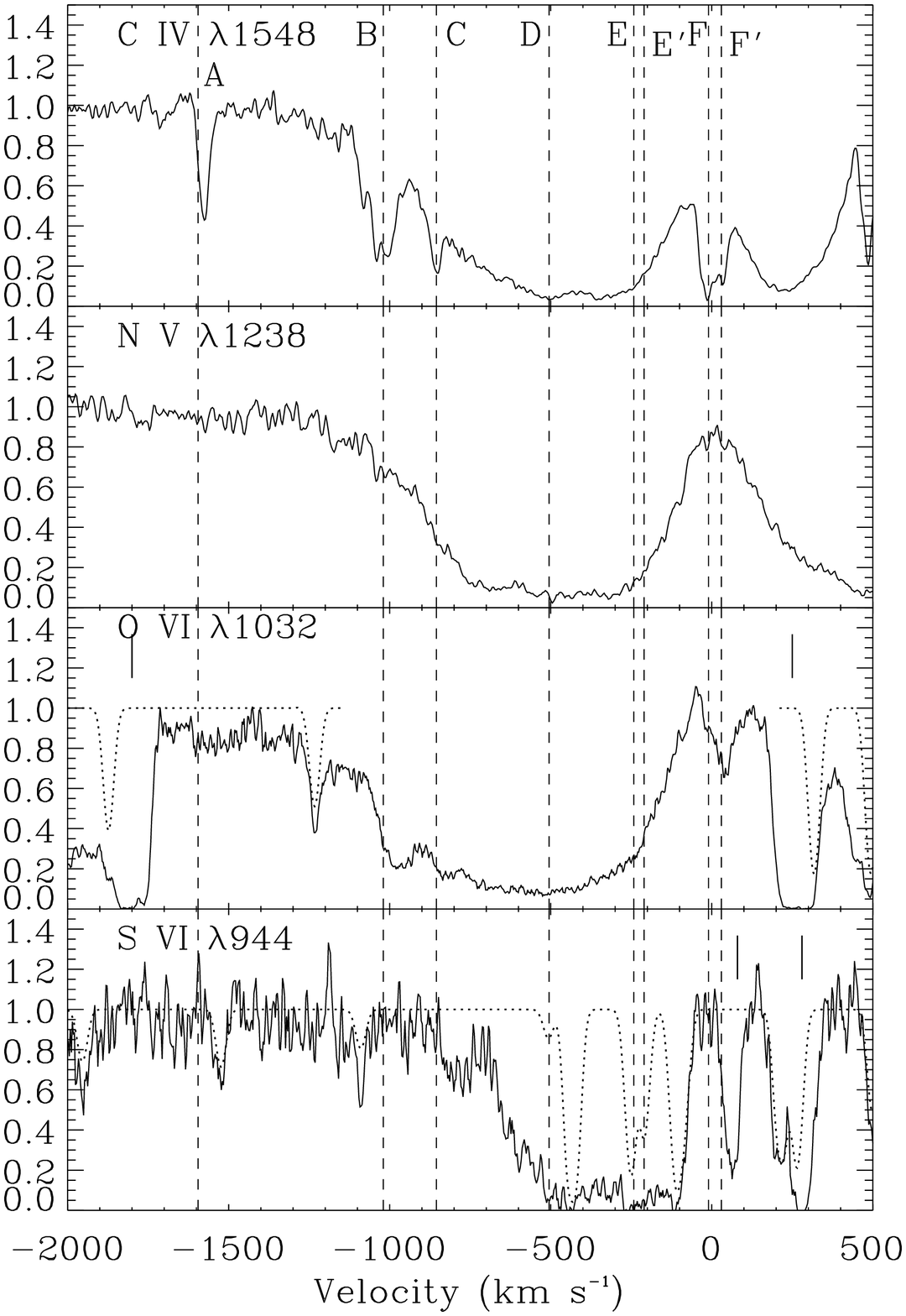}
\\Fig.~8a.
\end{figure}

\addtocounter{figure}{-1}

\vfill\eject
\begin{figure}
\epsscale{0.6}
\plotone{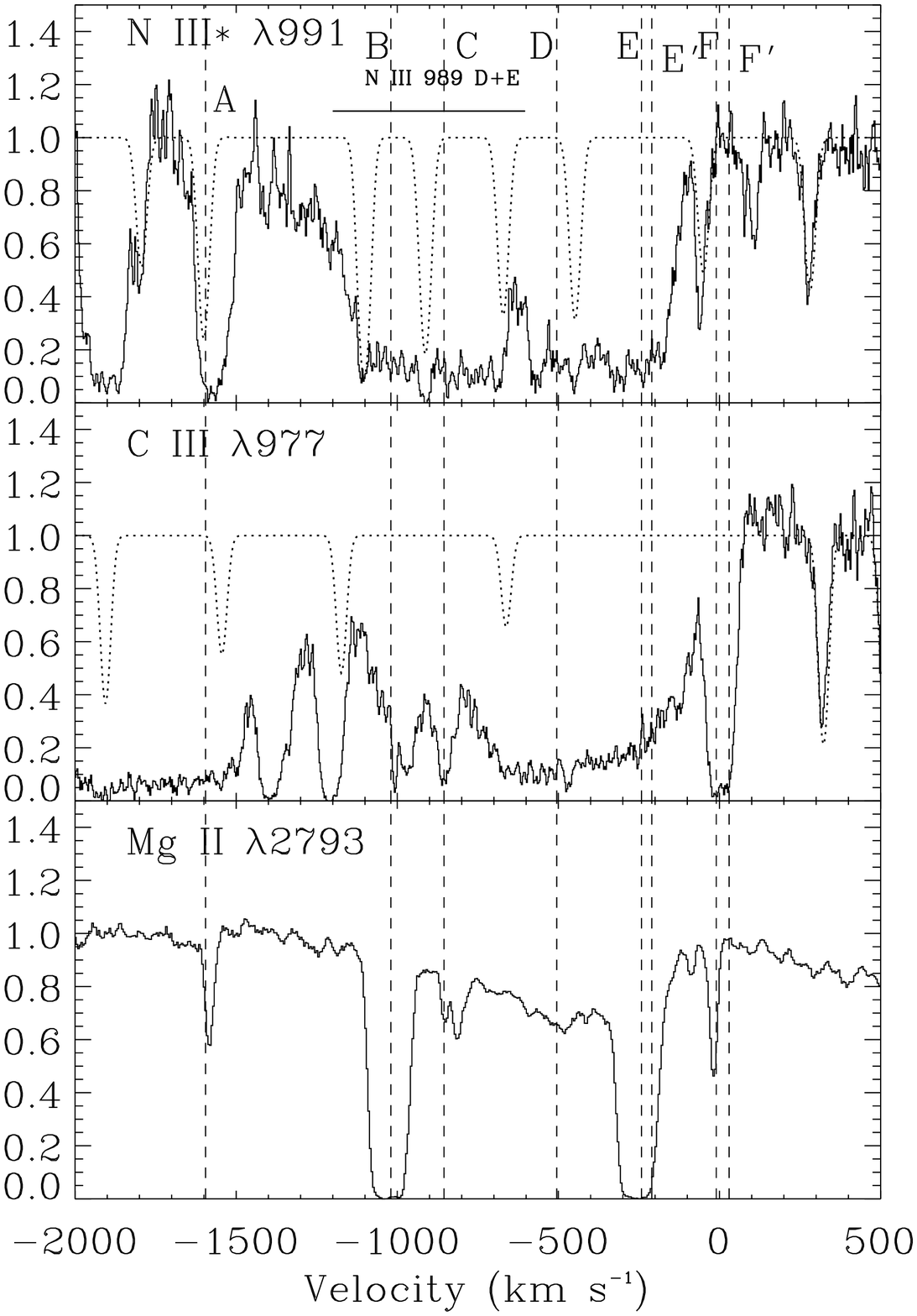}
\\Fig.~8b.
\end{figure}
\addtocounter{figure}{-1}

\vfill\eject
\begin{figure}
\epsscale{0.6}
\plotone{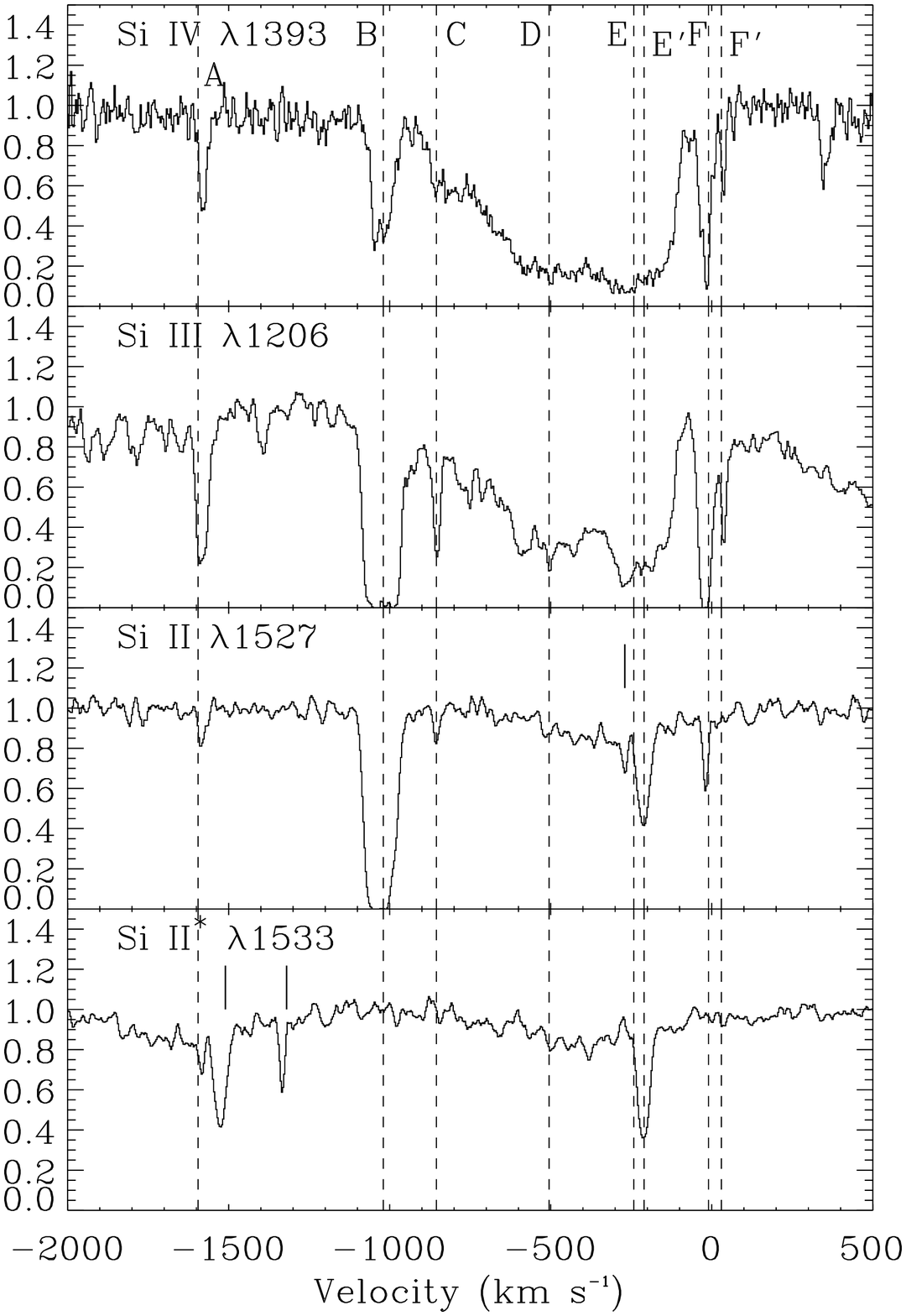}
\\Fig.~8c.
\end{figure}
\addtocounter{figure}{-1}

\vfill\eject
\begin{figure}
\epsscale{0.6}
\plotone{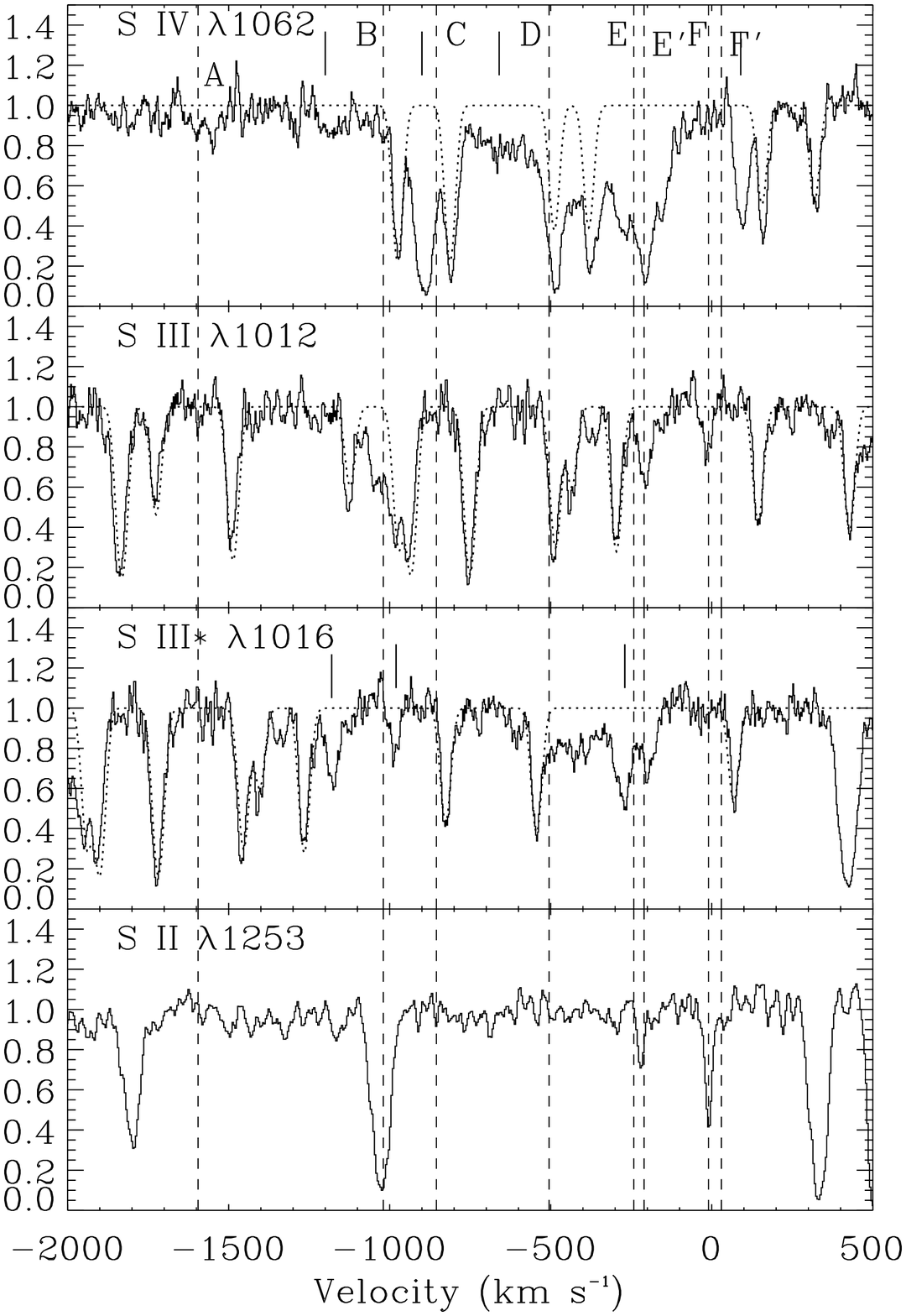}
\\Fig.~8d.
\end{figure}
\addtocounter{figure}{-1}

\vfill\eject
\begin{figure}
\epsscale{0.6}
\plotone{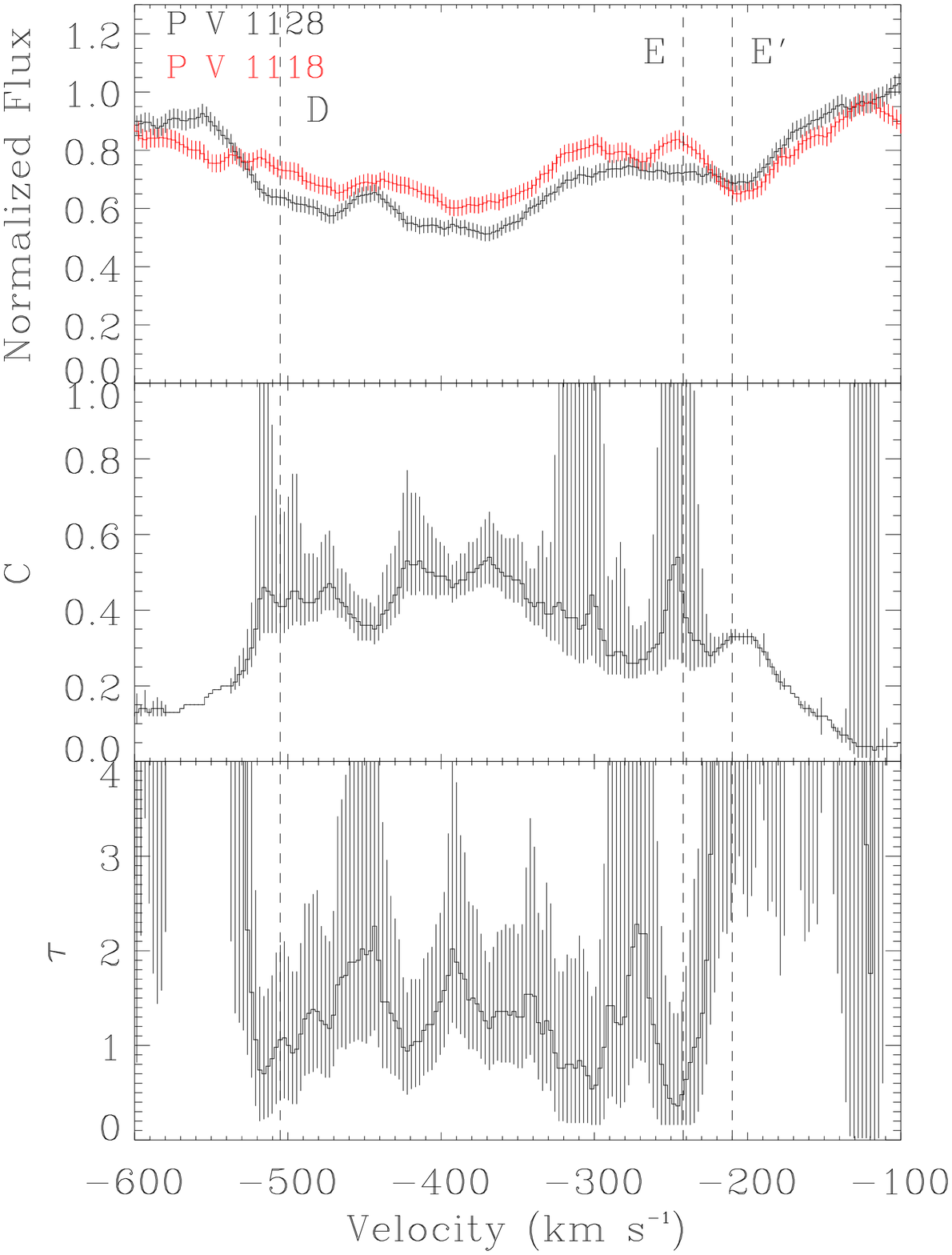}
\\Fig.~9.
\end{figure}

\vfill\eject
\begin{figure}
\epsscale{0.6}
\plotone{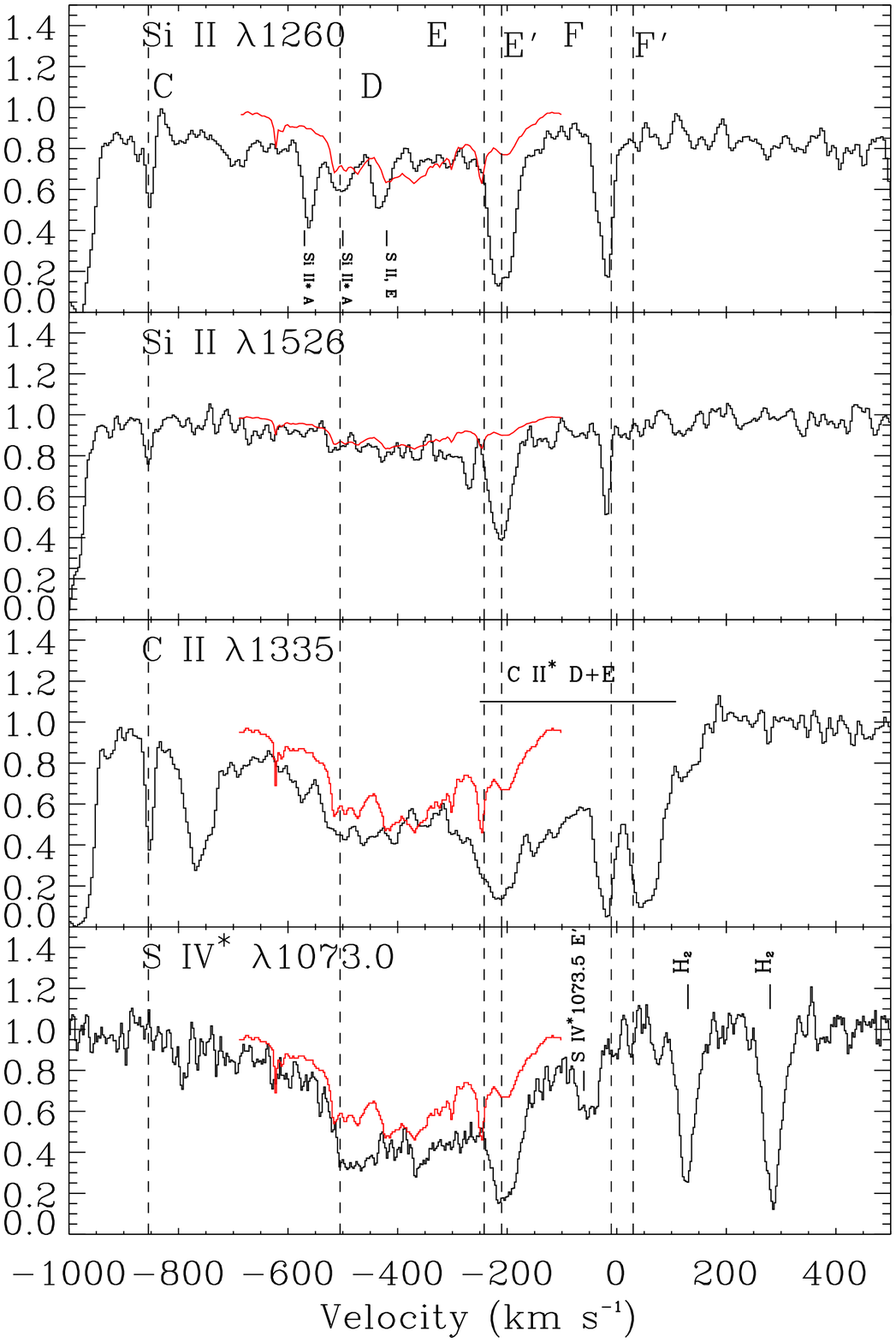}
\\Fig.~10.
\end{figure}

\vfill\eject
\begin{figure}
\plotone{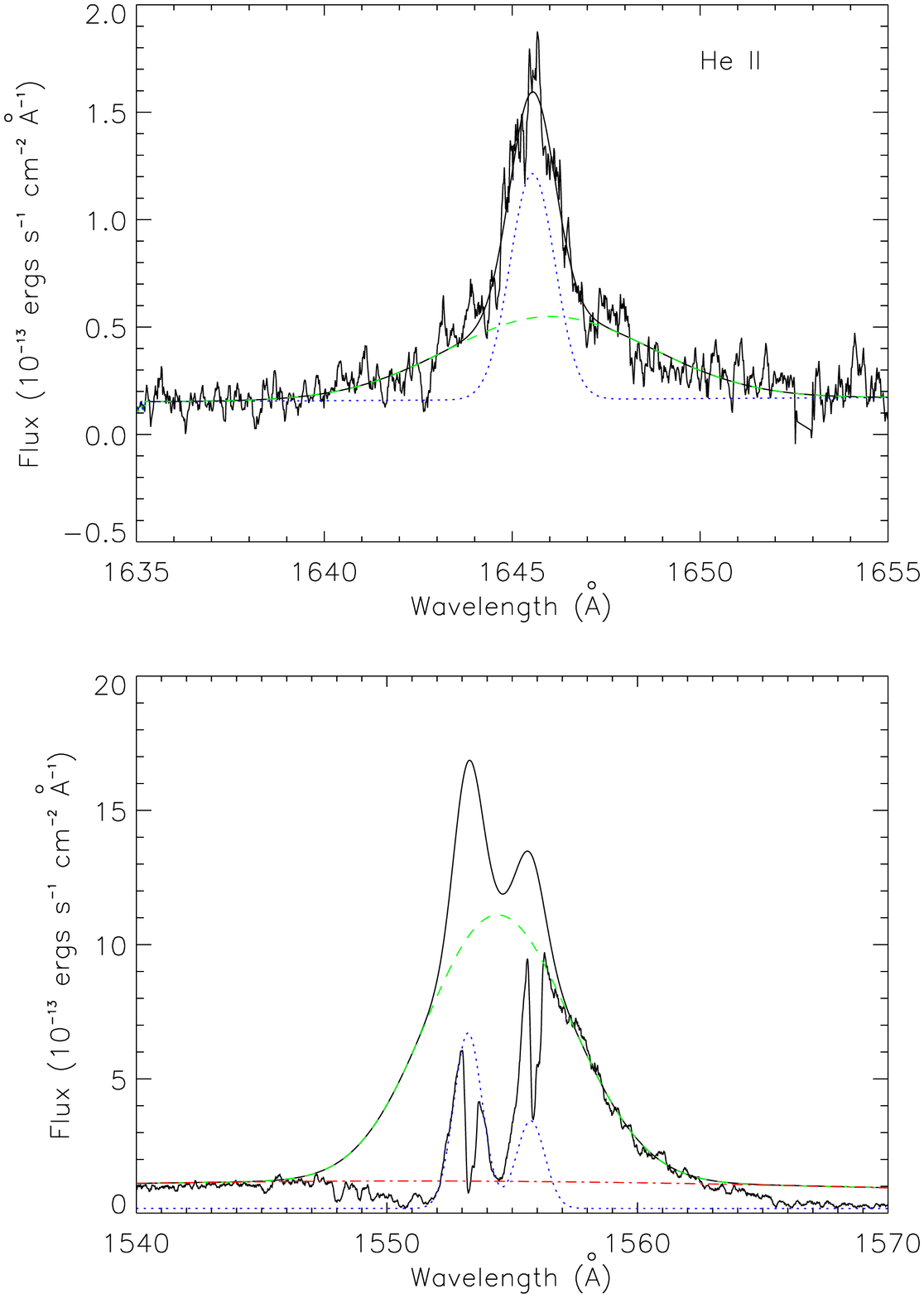}
\\Fig.~11.
\end{figure}

\vfill\eject
\begin{figure}
\epsscale{0.6}
\plotone{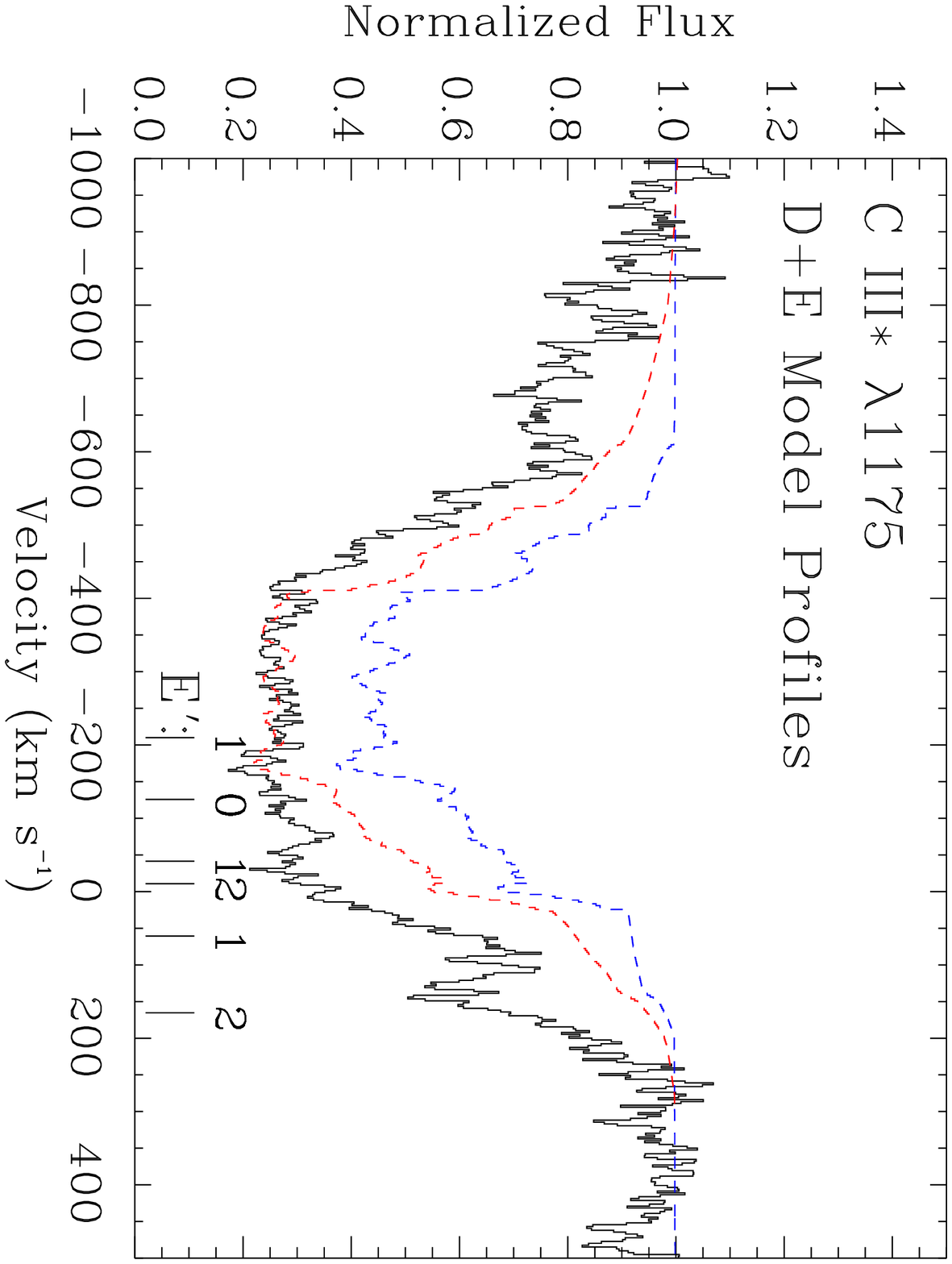}
\\Fig.~12.
\end{figure}

\end{document}